\documentclass[journal]{IEEEtran}
\usepackage{amsmath}
\usepackage{amssymb}
\usepackage{graphicx}
\usepackage{subfigure}
\usepackage{epsfig}
\usepackage{booktabs}
\usepackage{balance}
\usepackage{color}
\usepackage{graphicx,amssymb,amstext,amsmath}
\usepackage{leftidx}
\usepackage{cite}
\usepackage{microtype}
\usepackage{mathtools}
\usepackage[ruled,vlined,lined,ruled,linesnumbered]{algorithm2e}
\usepackage{float}
\usepackage{algorithmic}
\usepackage{hyperref}
\begin{document}

\title{Joint Resource Allocation and Cache Placement for Location-Aware Multi-User Mobile Edge Computing}
\author{Jiechen Chen, Hong Xing, \IEEEmembership{Member,~IEEE}, Xiaohui Lin,\\
	Arumugam Nallanathan, \IEEEmembership{Fellow,~IEEE}, and Suzhi Bi, \IEEEmembership{Senior Member,~IEEE}
\thanks{Part of this paper has been presented at the IEEE International Conference on Communications (ICC), June, 2020 \cite{my}. 
}
	\thanks{J. Chen, X. Lin and S. Bi are with the College of Electronics and Information Engineering, Shenzhen University, Shenzhen 518060, China (e-mails: chenjiechen2018@email.szu.edu.cn, \{xhlin, ~bsz\}@szu.edu.cn). S. Bi is also with Peng Cheng Laboratory, Shenzhen 518066, China.} 
	\thanks{H. Xing is with Internet of Things Thrust, The Hong Kong University of Science and Technology (Guangzhou), Guangzhou 511400, China (e-mail: hongxing@ust.hk). H. Xing is also with the Department of Electronic and Computer Engineering, The Hong Kong University of Science and  Technology, Hong Kong SAR, China.}
	\thanks{A. Nallanathan is with the School of Electronic Engineering and Computer Science, Queen Mary University of London, London  E1 4NS, U.K. (e-mail: nallanathan@ieee.org).}
}
\maketitle

\vspace{-10pt}
\maketitle

\newtheorem{definition}{\underline{Definition}}[section]
\newtheorem{fact}{Fact}
\newtheorem{assumption}{Assumption}
\newtheorem{theorem}{\underline{Theorem}}[section]
\newtheorem{lemma}{\underline{Lemma}}[section]
\newtheorem{proposition}{\underline{Proposition}}[section]
\newtheorem{corollary}[proposition]{\underline{Corollary}}
\newtheorem{example}{\underline{Example}}[section]
\newtheorem{remark}{\underline{Remark}}[section]
\newcommand{\mv}[1]{\mbox{\boldmath{$ #1 $}}}
\newcommand{\mb}[1]{\mathbb{#1}}
\newcommand{\Myfrac}[2]{\ensuremath{#1\mathord{\left/\right.\kern-\nulldelimiterspace}#2}}
\newcommand\Perms[2]{\tensor[^{#2}]P{_{#1}}}
\begin{abstract}
With the growing demand for latency-critical and computation-intensive Internet of Things (IoT) services, the IoT-oriented network architecture, {\it mobile edge computing (MEC)}, has emerged as a promising technique to reinforce the computation capability of the resource-constrained IoT devices. To exploit the cloud-like functions at the network edge, service caching has been implemented to reuse the computation task input/output data, thus effectively reducing the delay incurred by data retransmissions and repeated execution of the same task. In a multiuser cache-assisted MEC system, users' preferences for different types of services, possibly dependent on their locations, play an important role in joint design of communication, computation and service caching. In this paper, we consider multiple representative locations, where users at the same location share the same preference profile for a given set of services. Specifically, by exploiting the location-aware users' preference profiles, we propose joint optimization of the binary cache placement, the edge computation resource and the bandwidth allocation to minimize the expected sum-energy consumption, subject to the bandwidth and the computation limitations as well as the service latency constraints. To effectively solve the mixed-integer non-convex problem, we propose a deep learning (DL)-based offline cache placement scheme using a novel stochastic quantization based discrete-action generation method. The proposed hybrid learning framework advocates both benefits from the model-free DL approach and the model-based optimization. The simulations verify that the proposed DL-based scheme saves roughly 33\% and 6.69\% of energy consumption compared with the greedy caching and the popular caching, respectively, while achieving up to 99.01\% of the optimal performance. 
\end{abstract}

\begin{IEEEkeywords}
Mobile-edge computing, service caching, resource allocation, deep learning.
\end{IEEEkeywords}
\vspace{-10pt}
\section{Introduction}
The advent of the Internet of things (IoT) technologies has aroused the proliferation of new applications featuring intensive and real-time computation, such as virtual reality (VR), augmented reality (AR), online gaming, and autonomous driving, etc. \cite{2016:M}. However, fusion of data and service for these emerging types of applications tends to be prohibitive on front-end IoT devices such as various types of sensors, wearable devices and smart phones/tablets equipped with limited processing, storage and battery hardware. Meanwhile, to meet the demand for computation-intensive and latency-critical IoT services, these front-end IoT devices need to offload computation workload to cloud-like processing facilities for high-performance computing. However, the (ultra-)low latency requirement of these applications will not tolerate the long end-to-end latency due to transmission over fronthaul like in conventional cloud radio network (CRAN) \cite{2015:l}. To address this issue, fog radio access network (FRAN) has been proposed to provide cloud functionality down to the proximity of the IoT devices at the network edge, thus enabling $\emph{fog computing}$ or $\emph{mobile edge computing (MEC)}$ \cite{2017:Y}. Thanks to this IoT-oriented network architecture, the IoT devices are able to get fast response to their service requests and save large amount of energy \cite{2018:B}.

One of the mainstreams on MEC in the literature has centered on joint optimization of communication and computation resource allocation \cite{2017:Y,2018:B,2019:H,2016:L,2018:b,2018:ff,2019:M,2020:f,2018:x}, such as transmit power and bandwidth to achieve energy-efficient and low-latency computation. However, the above line of work has not incorporated another dimension of design, \emph{service caching} (or \emph{task caching}). Service caching refers to fetching \emph{a priori} task input data, program files or task results of frequently demanded computation services at edge servers or mobile devices, thus alleviating transmission and execution burden for (partially) repeated request in the future. As a result, service caching further unleashes potential of MEC in terms of energy efficiency and low latency \cite{2019:bi, 2019:h, 2017:cui, 2020:sun, 2020:w, 2020:a}. Note that content-oriented caching has been a well-investigated topic aimed for improving user-perceived quality of experience by reducing network congestion, especially for video content delivery. \textcolor{black}{For example, \cite{2016:T} studied a freshness-aware content refreshing scheme to balance the service delay and content freshness.} However, there are several different aspects lying between content-oriented and computation-oriented caching. i) Compared with content-oriented caching that mainly fetches data over backhaul, service caching takes place in the shared wireless medium, and is thus more vulnerable to channel hostilities such as channel noise, fading and mutual interference \cite{2017:cui, 2020:sun}. ii) Computation-oriented tasks are usually context-aware and customized to real-time data generated locally at mobile devices, and therefore the validity of task input/output data and/or program files may last relatively shorter than the content, incurring significant overhead due to service-caching redeployment \cite{2019:bi, 2019:h, 2017:cui}.
\subsection{Related Work}
There exists rich literature on joint design of communication and computation resource \cite{2019:H, 2016:L, 2018:b, 2018:ff, 2019:M}. The authors in \cite{2016:L} and \cite{2019:H} considered device-to-device (D2D)-enabled multi-helper MEC systems with multiple tasks, and jointly optimized task offloading and resource allocation to minimize total energy consumption and latency, respectively. In a wireless energy harvesting setup, \cite{2018:b} jointly optimized task offloading decisions and resource allocation assuming binary offloading to maximize the weighted sum computation rate of all users. The total energy consumption was minimized in \cite{2018:ff} by joint optimization of resource allocation, partial task offloading policies and energy transmit beamforming at the access point. \cite{2019:M} exploited spatial correlation among tasks of VR users located closely in proximity to reduce both the uplink and downlink traffic load in a multi-BS multi-user cloud computing setup.

On another front, there are also prior work that investigated performance gain brought by service caching. For example, \cite{2019:bi} studied a single-user cache-assisted MEC system with dependent tasks, and minimized the average computation latency and energy consumption considering the coupling effect of service cache placement and computation offloading decisions. \cite{2019:h} exploited temporal correlation among sequential task arrivals at a single user to enable proactive caching of partial task results, therefore reducing the total computation energy over a finite time horizon. In addition, \cite{2020:sun} assumed that the edge server has the input and the output data of all the computation tasks in a multi-user MEC system. Accordingly, it jointly optimized the local caching decisions of task input and/or output data and computing mode of mobile devices to minimize the transmission bandwidth.

Despite of these previous arts on the integrated design of communication, computation, and caching (3C), some assumed fixed service demand \cite{2019:bi}, which may not be valid in practice, as users normally have random request over different types of services. Although some of the work considered how the MEC users' preference profiles affect cache placement designs, they did not jointly optimize communication and computation resources along with cache placement. For example, \cite{2016:T} exploited content caching by designing an algorithm to learn users' preference profile, but it cannot apply to service caching with computation-oriented service requirements. For example, in a VR-assisted museum tour guide system, tourists at different display stands often make requests for context-aware introduction of different displays and are then served by videos played on their individual tour-guide handsets \cite{2018:Ad}. In this example, the task-input data can be the users' individual field of vision and the environment parameters at the display stands etc., and the (cached) task results is the video clips. For one realization of service delivery as illustrated by Fig. 1, both users at location $3$ and one user at location $2$ request a popular service $s_3$, but only the user at location $2$ who has the best channel condition among these three users needs to offload the task-input data of $s_3$ to the BS. Meanwhile, one user at location $1$, the edge of the service range, demands a computation-expensive service $s_1$. In this situation, it becomes a natural question to ask whether we place cache for service $s_3$ or $s_1$ subject to limited cache capacity at the BS, in order to minimize the expected sum energy consumption of the users with respect to their dynamic service requests.

Furthermore, cache placement design usually involves mixed-integer non-linear programming (MINLP) due to binary caching decision variables, which lacks efficient algorithm to solve in general. There are some existing works that adopted reinforcement learning (RL)-based methods to solve the MINLP in MEC settings. For example, \cite{2019:x} employed double DQN to optimize the offloading policy in an MEC system. \cite{2019:y} and \cite{2020:y} adopted actor-critic DRL and deep deterministic policy gradient (DDPG), respectively, to optimize caching strategies. \cite{2020:j} proposed a DDPG-based actor module to obtain resource allocation and a DQN-based critic module to select the best offloading decision. However, in these scenarios, value-based methods are computationally expensive due to the large and hybrid integer-continuous action space. In addition, policy-based methods often suffer from slow convergence, especially when the critic  module fails to produce an accurate and stable approximation of the value function early enough \cite{bi}. \cite{liang} proposed a hybrid learning-optimization framework and demonstrated its high efficiency in tackling the binary offloading problems in MEC networks. However, the effectiveness for this method to jointly optimize the binary cache placement as well as the (continuous) bandwidth (BW) and computation resource allocation is unknown. 
\begin{figure}[htp]
	\centering
	\includegraphics[width=3.45in]{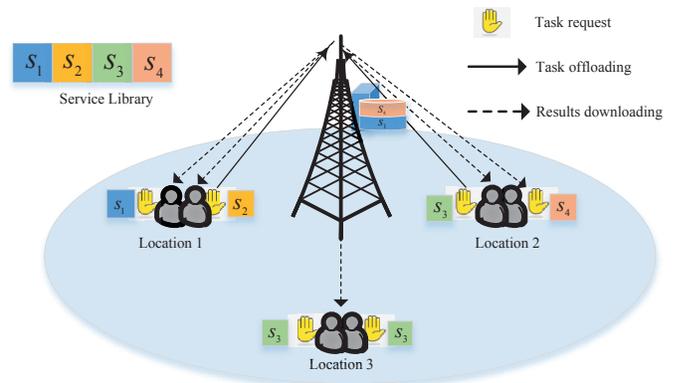}
	\caption{An illustration of a location-aware multi-user MEC system.}
	\label{model}
\end{figure} 
\vspace{-4ex}
\subsection{Contributions}
To tackle the above challenges, in this paper, we consider a multi-user MEC system equipped with narrow-band wireless communication facilities, where users request delay-sensitive computation services based on their location-dependent preferences. The users are then clustered by a fixed number of locations, and each location is representative of the users who share the same service demand profile. Then, among the users that request the same type of service, any user at a  location of the best channel condition will be selected to offload the task; and the BS will meet the demand by multicasting the computation results of the service at a rate that ensures successful delivery at all these locations. We aim for minimizing the expected weighted-sum energy consumption with respect to the users' preference profiles by joint optimization of cache placement, edge computation resources, and BW allocations. This problem is subject to instantaneous service deadline constraints, the maximum caching and computation capacities at the BS, as well as the BW constraints for data transmission. To effectively obtain the binary caching decisions, we propose a deep learning (DL) based offline cache placement scheme to solve the one-shot MINLP. The main contributions are summarized as follows.
\begin{itemize}
	\item We consider  multiple representative locations to simplify the problem of multi-user resource allocation and cache placement. This formulation necessitates only the channel state information (CSI) between several locations and the BS, thus facilitates the communications design, and also make the complexity of the problem scale with the number of types of services. 
	\item To obtain an optimal solution to the resource allocation problem given cache placement, we leverage Lagrangian dual decomposition method to solve the problem. The optimization framework used in this stage forms an essential module for the proposed DL-based cache placement policies.
	\item To solve the MINLP that are adaptive to dynamics of channels and service requests, we propose a DL-based hybrid (offline) learning framework to attain suboptimal caching decisions which advocates both benefits from the model-free DL approach and the model-based optimization. Specifically, the model-free DL module maps the input of channel and task state information to multiple binary caching decisions via a deep neural network (DNN), and the model-based optimization module selects the set of caching decisions that yields the best energy performance by solving multiple resource-allocation problems given the caching decisions. Thanks to the model-based module that provides accurate evaluation of the candidate caching decisions, the proposed hybrid learning framework simplifies the joint optimization problem to a classification problem, and also admits faster convergence than a fully model-free DL approach.
	\item We adopt a novel stochastic quantization based discrete-action generation scheme that samples candidate caching decisions from Bernoulli distribution based on the current model outputs, improving diversity in exploring the optimal caching decisions. 
	\item In special cases when users in one location only request one specified type of service, by exploiting the structure of the optimal solution, we can recast the original problem into a integer linear programming (ILP), which admits low-complexity solution using off-the-shelf software toolboxes, and attain effective suboptimal caching decisions.    
	\item Numerical results show the distinguishing performance gain brought by service caching in general and the efficacy of the proposed stochastic quantization based offline cache placement, by comparison with other benchmarks. 
\end{itemize}

The remainder of this paper is organized as follows. The multi-user MEC system model is presented in Section \ref{sec:System Model}. Section \ref{problem formulation} formulates the expected weighted-sum energy minimization problem. The jointly optimal solution for communication and computation resource allocation to the problem is investigated in Section \ref{sec:Optimal Solution}, with DL-based offline cache placement proposed in Section \ref{sec:Proposed solution}. The special case is studied in Section \ref{special}. Numerical results are provided in Section \ref{sec:Numerical Results}. Finally, Section \ref{sec:Conclusion} concludes the paper.

{\it Notation}---The superscript $([ \cdot ])^T$ represents the transpose of vectors. \(\mathbb{R}^{M\times N}\) stands for the sets of real matrices of dimension \(M\times N\). The cardinality of a set is represented by \(\vert\cdot\vert\). ${\rm Exp}(\lambda)$ denotes the exponential distribution with rate parameter $\lambda$. $\|\cdot\|$ denotes the Euclidean norm of a vector. In addition, \(\Pr(\cdot)\) means the probability of a random event.

\section{System Model}\label{sec:System Model}
As shown in Fig.~1, consider a MEC system which consists of a base station (BS) equipped with an edge server of caching facilities and multiple user-ends (UEs) grouped into $K$ different locations, denoted by $\mathcal{K}=\{1,\ldots,K\}$. Assume that there is a finite library of computation-intensive services denoted by $\mathcal{S}=\{s_1,\ldots,s_L\}$. Each computation service is characterized by a three-item tuple $(C_l,Q_l,R_l)$, $l=1,...,L$. Here, $C_l$ denotes an application-specific computation requirement of the $l$-th service (in CPU cycles per bit); $Q_l$ and $R_l$ denote the input and output data sizes of the computation service (in bits), respectively. The BS and the UEs are all equipped with a single antenna. We assume that all services can only be executed by the applications installed on the edge server due to the UEs' low-end IoT devices (such as sensors) with limited computation capacity and power supply \cite{2020:K} \cite{2019:G}. Since the task-input data corresponding to a specific service is assumed to remain unchanged for a period of time as a result of the slow-changing environment dynamics, the associated task-output data keeps the same and can thus be generated and cached {\it a priori} at the BS for reuse during the considered period of time.

Under this setup, UEs at different locations $k\in\mathcal{K}$ make requests for computation services, and then the BS aggregates the requests and matches them with its cache placement. If the task results for a required service is proactively cached at the BS, the BS will broadcast the cached task results to the target UEs. Otherwise, the UEs must first offload the task-input data to the BS, and then the BS performs necessary computation before broadcasting the task results to all UEs demanding this service. In one round of end-to-end computation service delivery, the MEC workflow consists of three phases: 1) task offloading, 2) task computation and 3) results downloading. In the first phase, the representative UEs (who have the best channel conditions among all those who request the same (uncached) service) offload their respective task-input data to the BS. In the second phase, the BS executes the (uncached) service. In the third phase, the BS multicasts the task results of different services to the UEs. We will elaborate on these phases in the following subsections.
\subsection{Location-Aware Task Computation Model}
We consider one-shot task requests raised from users at different locations. Specifically, we assume that UEs in one location follow the same task request distributions. We define by a matrix  $\mv A\in\mathbb{R}^{L\times K}$ the tasks' request state, whose $(l,k)$-th entry, $A_{l,k}\in\left\{0,1\right\}, s_l\in\mathcal{S}, k\in\mathcal{K}$, is given by
\begin{align}
A_{l,k}=\begin{cases}
1, & \mbox{if there is a UE at location $k$ requesting} \\ 
& \mbox{computation service $s_l$},\\
0, & \mbox{otherwise}.
\end{cases}\label{A} 
\end{align}
Also, we denote the fixed probability mass function (PMF) for a UE at location $k\in\mathcal{K}$ to demand service over $\mathcal{S}$ by $P_{l,k}=$ \(\Pr(A_{l,k}=1)\), such that \(\sum_{s_l\in S}P_{l,k}=1\), \(\forall k\in \mathcal{K}\).  Note that in general $P_{l,k}\neq P_{l,k^\prime}$ for any $k,\, k^\prime\in\mathcal{K}$, \(k\neq k^\prime\).

The BS can proactively cache the computation results of some services to eliminate their real-time execution delay. We define cache placement decisions against service $s_l\in\mathcal{S}$, by an indicator function as follows. 
\begin{align}
I_l=\begin{cases}
1, & \mbox{if the results of $s_l$  are cached at the BS,} \\ 
0, & \mbox{otherwise}.
\end{cases}
\end{align} 

The maximum caching capacity equipped on the BS is assumed to be $S$ (in bits), i.e.,\footnote{We assume a type of on-chip caching facilities that incurs negligible accessing delay.}
\begin{align}
\sum_{l=1}^LI_lR_l \leq S. \label{C:caching capacity constraint}
\end{align}
Note that we assume $\sum_{l=1}^{L}R_l > S$ by default, since the results of all types of services can all be cached otherwise, which reduced to a trivial solution of $I_l=1, \forall s_l\in \mathcal{S}$.

We define by $\mathcal{K}_l=\{k\in\mathcal{K}|A_{l,k}=1\}$ the set of locations where UEs demand service \(s_l\in\mathcal{S}\). The BS needs to provide the computation result of the $l$-th service if and only if $|\mathcal{K}_l|\geq 1$, $\forall s_l \in S$. We adopt a commonly used computation model \cite{2019:h}, in which the total number of CPU cycles required for performing one computation task is linearly proportioned to its task input bit length. As a result, the total number of CPU cycles required for the $l$-th task is given by $C_lQ_l$. We assume a multi-core CPU architecture at the edge server, so that each offloaded task is processed by a different core\cite{2018:b}. Thanks to dynamic voltage and frequency scaling techniques (DVFS) \cite{2019:H}, we denote the variable computation frequency (in cycles per second) and the incurred delay for processing the $l$-th task as $f_l$ and $t_l^{\rm c}$, which are related by 
\begin{align}
t_l^{\rm c}=\begin{cases}
\frac{C_lQ_l}{f_l}(1-I_l), & \mbox{if $|\mathcal{K}_l| \geq 1$}, \\ 
0, & \mbox{otherwise}.\label{frequency}
\end{cases}
\end{align}
Notice that we simply set $t_l^{\rm c}=f_l=0$ for service $s_l$ with $|\mathcal{K}_l|=0$. Equation \eqref{frequency} implies that the BS does not need to recompute the cached computation result with $I_l =1$. A common maximum computation frequency constraints are applied to all the computation cores, i.e.,\footnote{If a MEC setup also imposes the sum computation capacity constraint, i.e.,  $\sum_{s_l\in\mathcal{S}}f_0^{\rm max}(1-I_l)\leq f^{\rm max}$, which is equivalent to add a linear constraint on the cache placement decisions, the proposed solution of this paper is still applicable subject to minor changes.}
\begin{align}
f_l\le f_0^{\max}, \; \forall s_l\in\mathcal{S}. \label{fre}
\end{align}
Accordingly, the energy consumed by the BS for executing service $s_l$ is expressed as \cite{2019:h}
\begin{align}
E_l^{\rm c}=\begin{cases}
\kappa_0\frac{(C_lQ_l)^3}{(t_l^{\rm c})^2}(1-I_l), & \mbox{if $|\mathcal{K}_l| \geq 1$}, \\ 
0, & \mbox{otherwise},
\end{cases}
\end{align} 
where \(\kappa_0\) is a constant denoting the effective capacitance coefficient of the server chip architecture. The expected computation energy consumed by the BS for executing task $s_l\in\mathcal{S}$ w.r.t the users' request for task $s_l\in\mathcal{S}$ is thus given by
\begin{multline}
\mathbb{E}[E_l^{\rm c}]=0\times \Pr(|\mathcal{K}_l|=0) +\frac{\kappa_0(C_lQ_l)^3}{(t_l^{\rm c})^2}(1-I_l)\\\times (1-\Pr(|\mathcal{K}_l|=0)). 
\end{multline}
As \(|\mathcal{K}_l|=0\) means that no UE in any location requests service $s_l$, $\Pr(|\mathcal{K}_l|=0)$ is expressed as
\begin{align}
\Pr(|\mathcal{K}_l|=0) =\Pr(\bigcap\limits_{k=1}^K A_{l,k}=0)=\prod_{k\in\mathcal{K}}(1-P_{l,k}).
\end{align}
Hence, the expected total computation energy for executing all the request tasks is 
\begin{align}
E^{\rm c}&=\sum_{l=1}^L\mathbb{E}[E_l^{\rm c}] \notag\\
&=\sum_{l=1}^L\frac{\kappa_0(C_lQ_l)^3}{(t_l^c)^2}(1-I_l)(1-\prod_{k\in\mathcal{K}}(1-P_{l,k})).\label{eq:BS computation energy}
\end{align} 

\subsection{Location-Aware Communication Model}
In this subsection, we introduce the communication models for task offloading and results downloading. We assume that task offloading and result downloading phases are assigned with separate narrow bands with a total BW of $B$ (in Hz), respectively. The transmissions for different services are performed over orthogonal bandwidth via frequency-division multiple access (FDMA). We define the BW allocated to service $s_l\in\mathcal{S}$ for task offloading (results downloading) by $B_l^{\rm off}=\alpha_l^{\rm off}B$ ($B_l^{\rm dl}=\alpha_l^{\rm dl}B$), where $\alpha_l^{\rm off} (\alpha_l^{\rm dl}) \in [0,1]$ is the proportion of the BW allocated to service $s_l$, such that $\sum_{s_l\in\mathcal{S}}B_l^{\rm off}=B$ ($\sum_{s_l\in\mathcal{S}}B_l^{\rm dl}=B$). In addition, we assume slow fading scenarios, where the wireless channels remain constant during a specified period (shorter than the channel coherence time), which is defined to be as long as several computation deadline. We also assume that UEs in one location are identical in their path-loss factors and small-scale fading{\color{black}\footnote{{\color{black}This simplified location-representation multi-user channel models can apply to practical scenarios, e.g., where the users at the same location request computation services by connecting to a common gate way and they are connected with the common gateway via dedicated fiber with negligible transmission cost \cite{2020jia}.}}}. We denote \( h_k^{\prime}\) and \( g_k^{\prime}\) as channel coefficients between location $k\in \mathcal{K}$ and the BS for task offloading and results downloading, respectively. We assume that \( h_k^\prime=\sqrt{A_0}(\Myfrac{d_0}{d_k})^{\Myfrac{\gamma}{2}} h_k\) $\big( g_k^\prime=\sqrt{A_0}(\Myfrac{d_0}{d_k})^{\Myfrac{\gamma}{2}} g_k\big)$, \(k\in\mathcal{K}\), consists of Rayleigh fading with $h_k$ $(g_k)\sim\mathcal{CN}(0, 1)$ and multiplicative path loss \(\sqrt{A_0}(d_0/d_k)^{\gamma/2}\), where \(A_0\) is the average channel power gain at reference distance $d_0$; $d_k$ is the distance between location $k$ and the BS, and $\gamma$ denotes the path loss exponent factor. Without loss of generality, we also assume descending orders for the normalized channel gains as $u_1\geq \dots\geq u_K$, where \(u_k=\Myfrac{\|h_k^\prime\|^2}{(N_0B)}\) is the normalized channel gains with $N_0$ being the power spectral density of the additive white Gaussian noise (AWGN). Besides, we assume  $v_{\pi(1)}\geq \dots\geq v_{\pi(K)}$, where \(v_{\pi{(k)}}=\Myfrac{\|g_{\pi{(k)}}^\prime\|^2}{(N_0B)}\) and $\pi(\cdot)$ denotes a permutation over $\mathcal{K}$.

1) {\bf Task Offloading.}
The achievable rate for offloading task $s_l\in\mathcal{S}$ from any user at location $k\in \mathcal{K}$ is given by
\begin{equation}
\begin{aligned}
r_{l,k}^{\rm{off}}=\alpha_l^{\rm{off}} B\log_{2}\left(1+{\frac{p_k^{\rm{off}} u_k}{\alpha_l^{\rm{off}}}}\right),
\end{aligned} \label{offloading rates}
\end{equation}
where $p_k^{\rm off}$ is the transmitting power at location $k$. The transmission latency due to offloading service $s_l\in\mathcal{S}$ from location $k\in\mathcal{K}$ is thus expressed as 
\begin{align}
t_{l,k}^{\rm off}=\frac{Q_l}{r_{l,k}^{\rm off}}(1-I_l). 
\end{align}

When $|\mathcal{K}_l|\geq1$ locations demand the same computation service $s_l\in\mathcal{S}$, we choose the location among $\mathcal{K}_l$ with the best (normalized) channel gain to perform task offloading so as to reduce the transmission latency and energy consumption. The energy consumed in offloading service $s_l\in\mathcal{S}$ from location $k\in\mathcal{K}_l$ is:
\begin{align}
E_{l,k}^{\rm off}=\begin{cases}
p_k^{\rm off}t_{l,k}^{\rm off}, & \mbox{if a UE from location $k$ performs} \\ 
& \mbox{task offloading of service $s_l$},\\
0, & \mbox{otherwise}.
\end{cases}
\end{align}

If a UE from location $k\in\mathcal{K}_l$ is selected to offload service $s_l$, no user demands service $s_l$ from any locations with larger channel gains to the BS than location $k$. As a result, the probability that an UE from location $k$ is selected to offload service $s_l\in \mathcal{S}$ is expressed as follows:
\begin{align}
P_{l,k}^{\rm off}=\begin{cases}
\Pr\big((\bigcap\limits_{j=1}^{k-1}A_{l,j}=0)\bigcap A_{l,k}=1\big), & \mbox{if $k>1$}, \\ 
\Pr(A_{l,1}=1), & \mbox{if $k=1$},
\end{cases}
\end{align}
which can be simplified as 
\begin{align}
P_{l,k}^{\rm off}=\begin{cases}
\prod_{j=1}^{k-1}(1-P_{l,j})P_{l,k}, & \mbox{if $k>1$}, \\ 
P_{l,1}, & \mbox{if $k=1$}.
\end{cases}
\end{align}
The corresponding expected energy for offloading service $s_l$ w.r.t task request distribution at location $k$ expressed as 
\begin{align}
\mathbb{E}[E_{l,k}^{\rm off}]=p_k^{\rm off}t_{l,k}^{\rm off}P_{l,k}^{\rm off}.
\end{align}
The total expected task offloading energy w.r.t demand at location $k\in\mathcal{K}$ is thus given by 
\begin{align}
E_k^{\rm off}=\sum_{l=1}^L\mathbb{E}[E_{k,l}^{\rm off}]=p_k^{\rm off}\sum_{l=1}^Lt_{l,k}^{\rm off}P_{l,k}^{\rm off}. \label{eoff}
\end{align}

\begin{figure}[htp]
	\centering
	\includegraphics[width=3.45in]{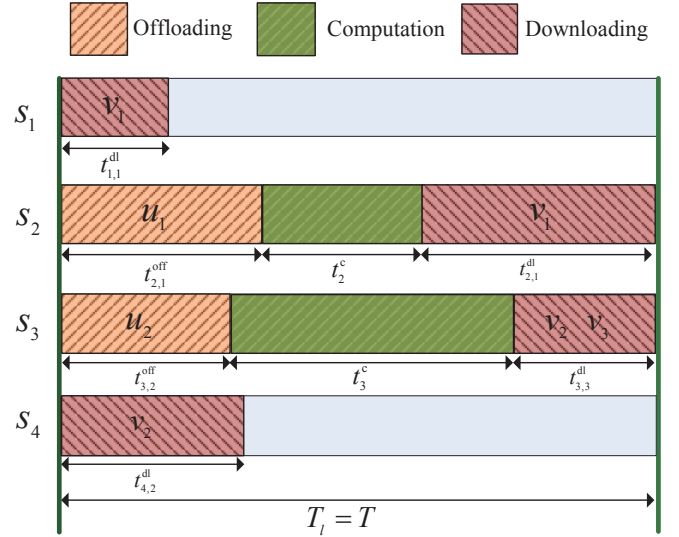}
	\caption{An illustration of the workflow for the MEC system shown in Fig. 1. Under the assumption of $u_1> u_2> u_3$ and $v_1>v_2>v_3$, for services $s_1$ and $s_4$, the first two phases can be skipped thanks to the cache placed {\it a priori}. The task-input data of service $s_3$ is offloaded by the user at location $2$, while the task-output data of service $s_3$ must be broadcast to accommodate the users at both location $2$ and location $3$. For service $s_2$, its task-related data transmission only occurs between a user at location $1$ and the BS.}
	\label{t}
\end{figure}

2) {\bf Results Downloading.}
After remote execution of service $s_l\in\mathcal{S}$, the BS transmits back the results to $\mathcal{K}_l$ by broadcasting, such that UEs from all these locations can download their desired results. Assuming that location $\pi(k)\in\mathcal{K}_l$ is of the worst normalized channel gain among the locations where service $s_l\in\mathcal{S}$ is requested, the transmission rate that the BS can successfully broadcast the results to UEs in $\mathcal{K}_l$ is expressed as  
\begin{align}
r_{l,\pi(k)}^{\rm dl} = \alpha_l^{\rm dl}B\log_2\left(1+\frac{p_l^{\rm dl}v_{\pi(k)}}{\alpha_l^{\rm dl}}\right), \label{downrate}
\end{align}
where $p_l^{\rm dl}$ is the transmitting power at the BS for service $s_l\in \mathcal{S}$.
The transmission latency caused by downloading the results of the $l$th service using rate $r^{\rm dl}_{l,\pi(k)}$ is $t_{l,\pi(k)}^{\rm dl}=\Myfrac{R_l}{r_{l,\pi(k)}^{\rm dl}}$. The energy consumed by the BS for broadcasting service $s_l$ is accordingly given by
\begin{align}
E_{l,\pi(k)}^{\rm dl}=\begin{cases}
p_l^{\rm dl}t_{l,\pi(k)}^{\rm dl}, & \mbox{if \(v_{\pi(k)}=\arg\min_{k\in\mathcal{K}_l}v_k\)}, \\ 
0, & \mbox{otherwise}. \label{Edownload}
\end{cases}
\end{align}

Equation \eqref{Edownload} implies that the UEs from all locations with smaller channel gains than location $\pi(k)$ (c.f. channel gains sorted  in descending order: as $v_{\pi(k+1)} \geq... \geq v_{\pi(K)}$) do not demand for service $s_l$. Accordingly, the probability of broadcasting service $s_l$'s results at the rate subject to location $\pi(k)$'s channel gain is given by 
\begin{align*}
&P_{l,\pi(k)}^{\rm dl} = \\
&\kern-4pt\left\{\kern-4pt
\begin{array}{ll}
{\Pr\big((\kern-2pt\bigcap\limits_{j=\pi(k+1)}^{\pi(K)}\kern-2ptA_{l,j}=0)\bigcap A_{l,\pi(k)}=1\big)}, &\text{if}\  \pi(k)<\pi(K), \\
{\Pr(A_{l,\pi(K)}=1)}, &\mbox{otherwise}, 
\end{array}  
\right.
\end {align*}
which can be simplified as 
\begin{align*}
&\kern-0.6inP_{l,\pi(k)}^{\rm dl} = \\
&\kern-0.6in\left\{\kern-4pt
\begin{array}{ll}
{\prod_{j=\pi(k+1)}^{\pi(K)}(1-P_{l,j})P_{l,\pi(k)}}, &\text{if}\ \pi(k)<\pi(K), \\
{P_{l,\pi(K)}}, &\mbox{otherwise}.  
\end{array}  
\right.
\end {align*}

The expected energy for broadcasting service $s_l$'s results w.r.t demand profile is
\begin{align}
E_l^{\rm dl}=\sum_{k=1}^K\mathbb{E}[E_{l,\pi(k)}^{\rm dl}]= \sum_{k=1}^K p_l^{\rm dl}t_{l,\pi(k)}^{\rm dl}P_{l,\pi(k)}^{\rm dl}.
\end{align}
The total expected transmission energy consumption at the BS is thus given by 
\begin{align}
E^{\rm dl}=\sum_{l=1}^L\sum_{k=1}^K\mathbb{E}[E_{l,\pi(k)}^{\rm dl}]=\sum_{l=1}^L\sum_{k=1}^Kp_l^{\rm dl}t_{l,\pi(k)}^{\rm dl}P_{l,\pi(k)}^{\rm dl}. \label{downloading energy}
\end{align}

To sum up, we illustrate in Fig. 2 the end-to-end workflow of the considered multi-user MEC system.

\section{Problem Formulation} \label{problem formulation}
In this section, we formulate the energy minimization problem. The expected weighted-sum energy consumed by the BS ($E^{\rm c}$ and $E^{\rm dl}$) and all UEs ($E_k^{\rm off}$'s) are given by \(\beta_0(E^{\rm c}+E^{\rm dl})+\sum_{k=1}^K\beta_kE_k^{\rm off}\), where \(\beta_0\ge 0\), \(\beta_k\ge 0\), and $\beta_0+\sum_{k\in\mathcal{K}}\beta_k=1$, are normalized weighted factors. \textcolor{black}{For example, when $\beta_0 = 0$, the objective function reduces to the energy consumption of the users only, and thus our formulated minimization problem is of flexible design to accommodate any level of energy-consumption trade-offs between the BS and the users in practice by tuning these weighted factors.} The total latency for delivering service $s_l\in\mathcal{S}$, i.e., $t_{l,k}^{\rm off}+t_l^{\rm c}+t_{l,j}^{\rm dl}$, for all $s_l\in\mathcal{S}$ and $(k,j)\in\mathcal{K}_l\times\mathcal{K}_l$, is subject to an instantaneous deadline constraint $T_l$.

\begin{remark}
	The formulation can be modified to accommodate expected latency constraints by \(\mb{E}[t_{l,k}^{\rm off} + t_l^{\rm c} + t_{l,j}^{\rm dl}]\le T_l\), but we consider herein the latency-critical scenarios where the latency constraint for service \(s_l\) must hold for every possible combination of $(k,j)\in\mathcal{K}_l\times\mathcal{K}_l$, thus incurring higher energy consumption than the average latency constraints in general.
\end{remark}

By denoting \(\mv I=[I_1, \ldots, I_L]^T, \mv \alpha^{\rm off}=[\alpha_1^{\rm off}, \ldots, \mv \alpha_L^{\rm off}]^T\), \(\mv \alpha^{\rm dl}=\)\([\alpha_1^{\rm dl}, \ldots, \alpha_L^{\rm dl}]^T\), \(\mv t^{\rm c}=[t_1^{\rm c}, \ldots, t_L^{\rm c}]^T\), \(\mv t^{\rm off}_l=[t_{l,1}^{\rm off}, \ldots, t_{l,K}^{\rm off}]^T\) and $\mv t_l^{\rm dl}=[t_{l,\pi(1)}^{\rm dl}, \ldots, t_{l,\pi(K)}^{\rm dl}]^T$, $s_l\in\mathcal{S}$, the expected weighted-sum energy minimization problem is formulated as:
\begin{subequations}
	\begin{align}
	\mathrm{(P0)}:\!\!\!\!&\mathop{\mathtt{Minimize}}_{\mv I,\mv \alpha^{\rm off},\mv \alpha^{\rm dl}, \mv t^{\rm c},\atop \left\{\mv t_l^{\rm off}\right\}_{s_l\in\mathcal{S}}, \left\{\mv t_l^{\rm dl}\right\}_{s_l\in\mathcal{S}}}\!\!\!
	\beta_0\left(E^{\rm c}+E^{\rm dl}\right)+\sum_{k=1}^K\beta_k E_k^{\rm off}\notag\\
	&\mathtt{Subject \ to}~~\eqref{C:caching capacity constraint},\notag\\
	& t_{l,k}^{\rm off}+t_l^{\rm c}+t_{l,j}^{\rm dl} \leq T_l, \forall s_l\in\mathcal{S},\;\forall (k,j)\in\mathcal{K}_l\times\mathcal{K}_l, \label{time}\\
	&t_l^{\rm c} \geq \frac{C_lQ_l(1-I_l)}{f_0^{\max}},\forall s_l\in\mathcal{S}, \label{computing capacity}\\
	&\sum_{s_l\in\mathcal{S}} \alpha_l^{\rm off} \leq 1,\label{total upbandwidth}\; \\
	&\sum_{s_l\in\mathcal{S}}\alpha_l^{\rm dl} \leq 1, \label{total downbandwidth} \\
	&\alpha_l^{\rm{off}} B\log_{2}\left(1+{\frac{p_k^{\rm off} u_k}{\alpha_l^{\rm{off}}}}\right) \geq \frac{Q_l(1-I_l)}{t_{l,k}^{\rm off}}, \notag\\
	&\forall k\in\mathcal{K},\; \forall s_l\in\mathcal{S},\label{offloading rate}   \\
	&\alpha_l^{\rm{dl}} B\log_{2}\left(1+{\frac{p_l^{\rm{dl}} v_{\pi(k)}}{\alpha_l^{\rm{dl}}}}\right) \geq \frac{R_l}{t_{l,\pi(k)}^{\rm dl}}, \notag\\
	&\forall k\in\mathcal{K},\forall s_l\in\mathcal{S}, \label{downloading rate} \\
	&I_l=\left\{0,1\right\}, \alpha_l^{\rm off}\in[0,1],\alpha_l^{\rm dl}\in[0,1], \forall s_l\in\mathcal{S}.\label{caching decision}
	\end{align}
\end{subequations}

The constraints in \eqref{computing capacity} are obtained by plugging \eqref{frequency} into the maximum frequency constraints (c.f.~\eqref{fre}). Constraints \eqref{total upbandwidth} and \eqref{total downbandwidth} are communication BW constraints for task offloading and results downloading, respectively. It is also worth-noting that constraints \eqref{offloading rate} and \eqref{downloading rate} are the minimum transmission rate requirements (c.f.~\eqref{offloading rates} and \eqref{downrate}), which can be easily shown to be active when $\mathrm{(P0)}$ is optimally solved.

In addition, problem $\mathrm{(P0)}$ can be further simplified by merging some of its constraints as follows.
\begin{lemma}
	Problem $\mathrm{(P0)}$ can be equivalently transformed to the following problem:
	\begin{subequations}
		\begin{align}
		\mathrm{(P0^\prime)}:\!\!&~\mathop{\mathtt{Minimize}}_{\mv I,\mv \alpha^{\rm off},\mv \alpha^{\rm dl}, \mv t^{\rm c}, \atop \{\mv t_l^{\rm off}\}_{s_l\in\mathcal{S}}, \{\mv t_l^{\rm dl}\}_{s_l\in\mathcal{S}}}\!\!\!
		~~\beta_0\left(E^{\rm c}+E^{\rm dl}\right)+\sum_{k=1}^K\beta_k E_k^{\rm off}\notag\\
		&~\mathtt{Subject \ to}~~\eqref{C:caching capacity constraint}, \eqref{computing capacity}-\eqref{caching decision}\notag\\
		&~ t_{l,K}^{\rm off}+t_l^{\rm c}+t_{l,\pi(K)}^{\rm dl} \leq T_l, \forall s_l\in\mathcal{S}.  \label{times}
		\end{align}
	\end{subequations}
\end{lemma}

\emph{Proof:} Constraints \eqref{time} include all cases where the transmission and execution delay for any task should be within deadline $T$. Hence, if the worst case with the longest service latency satisfies the deadline constraint, i.e., $t_{l,K}^{\rm off}+t_l^{\rm c}+t_{l,\pi(K)}^{\rm dl} \leq T_l, \forall s_l\in\mathcal{S}$, so do all other cases.

\begin{figure*}[tp]
	{ \small\begin{multline}		
		~\mathcal{L}(\mv P, \mv D)=\beta_0\sum_{l=1}^L\kappa_0\frac{(C_lQ_l)^3}{(t_l^c)^2}(1-\bar{I_l})(1-\prod_{k=1}^K(1-P_{l,k}))+\beta_0\sum_{l=1}^L\sum_{k=1}^Kp_l^{\rm dl}t_{l,\pi(k)}^{\rm dl}P_{l,\pi(k)}^{\rm dl}+\sum_{k=1}^K\beta_kp_k^{\rm off}\sum_{l =1}^Lt_{l,k}^{\rm off}P_{l,k}^{\rm off}\\\!\!\!\!\!\!\!+\sum_{l=1}^L\mu_l(t_{l,K}^{\rm off}+t_l^{\rm c}+t_{l,\pi(K)}^{\rm dl}-T_l)+\sum_{l=1}^L\eta_l\bigg(\frac{C_lQ_l(1-I_l)}{f_0^{max}}-t_l^{\rm c}\bigg)+\sigma(\sum_{l=1}^L\alpha_l^{\rm off}-1)+\epsilon(\sum_{l=1}^L\alpha_l^{\rm dl}-1)\\+\sum_{k=1}^K\sum_{l=1}^L\omega_{l,k}\bigg(\frac{Q_l}{t_{l,k}^{\rm off}}-\alpha_l^{\rm dl}B\log_2\big(1+\frac{p_k^{\rm off}u_k}{\alpha_{l}^{\rm off}}\big)\bigg)+\sum_{k=1}^K\sum_{l=1}^L\gamma_{l,k}\bigg(\frac{R_l}{t^{\rm dl}_{l,\pi(k)}}-\alpha_l^{\rm dl}B\log_2\big(1+\frac{p_l^{\rm dl}v_{\pi(k)}}{\alpha_l^{\rm dl}}\big)\bigg). \label{lagrangian}
		\end{multline}
	}\hrulefill 
\end{figure*}
\begin{figure*}[tp]
	{\small\begin{multline}
\mathcal{L}^{\prime}(\mv P, \mv D)=\bigg(\beta_0\sum_{l=1}^L\kappa_0(1-\bar{I_l})\frac{(C_lQ_l)^3}{(t_l^c)^2}\big(1-\prod_{k=1}^K(1-P_{l,k})\big)+\sum_{l=1}^L\mu_lt_l^{\rm c}-\sum_{l=1}^L\eta_lt_l^{\rm c}\bigg)+\bigg(\beta_0\sum_{l=1}^L\sum_{k=1}^Kp_l^{\rm dl}t_{l,\pi(k)}^{\rm dl}P_{l,\pi(k)}^{\rm dl}\\+\sum_{l=1}^L\mu_lt_{l,\pi(K)}^{\rm dl}+\sum_{k=1}^K\sum_{l=1}^L\gamma_{l,k}\frac{R_l}{t_{l,\pi(k)}^{\rm dl}}\bigg)+\bigg(\sum_{k=1}^K\beta_kp_k^{\rm off}\sum_{l=1}^Lt_{l,k}^{\rm off}P_{l,k}^{\rm off}+\sum_{l=1}^L\mu_lt_{l,K}^{\rm off}+\sum_{k=1}^K\sum_{l=1}^L\omega_{l,k}\frac{Q_l}{t_{l,k}^{\rm off}}\bigg)+\\\bigg(\sigma\sum_{l=1}^L\alpha_l^{\rm off}-\sum_{k=1}^K\sum_{l=1}^L\omega_{l,k}\alpha_l^{\rm off}B\log_2\big(1+\frac{p_k^{\rm off}u_k}{\alpha_{l}^{\rm off}}\big)\bigg)+\bigg(\epsilon\sum_{l=1}^L\alpha_l^{\rm dl}-\sum_{k=1}^K\sum_{l=1}^L\gamma_{l,k}\alpha_l^{\rm dl}B\log_2\big(1+\frac{p_l^{\rm dl}v_{\pi(k)}}{\alpha_l^{\rm dl}}\big)\bigg). \label{lagrangian2}
		\end{multline}
		\hrulefill}
\end{figure*}

\section{Optimal Communication And Computation Resource Allocation} \label{sec:Optimal Solution}
In this section, we study the optimal solution to problem $\mathrm{(P0^\prime)}$. Since problem $\mathrm{(P0^\prime)}$ is a MINLP that is in general NP-hard, we solve $\mathrm{(P0^\prime)}$ by decomposing it into two-stage optimization problems: 1) BW and edge computing resource allocation problem with the caching decisions fixed as $\mv I=\bar{\mv I}$, denoted as $\mathrm{(P0^\prime\text{-}1)}$; and 2) cache placement problem $\mathrm{(P0^\prime\text{-}2)}$ to find the optimal caching decisions.  In this section, we focus on solving $\mathrm{(P0^\prime\text{-}1)}$.

It is easily verified that $\mathrm{(P0^\prime\text{-}1)}$ is a convex problem, {\color{black}(The left-hand side (LHS) of constraints \eqref{offloading rate} and \eqref{downloading rate} are perspective of concave functions, and therefore prove to be concave w.r.t. $\alpha_l^{\rm off}$ and $\alpha_l^{\rm dl}$, respectively.)} and also satisfies Slater's condition. Hence, we leverage Lagrangian dual decomposition method to solve problem $\mathrm{(P0^\prime\text{-}1)}$ with strong duality guaranteed \cite{convex}.  

By denoting the primal-variable tuple and dual-variable tuple as $\mv P=(\mv\alpha^{\rm off}, \mv\alpha^{\rm dl},\{\mv t_l^{\rm dl}\}, \{\mv t_l^{\rm off}\},\mv t^{\rm c})$ and $\mv D=(\mv\mu, \mv\eta,\mv\omega, \mv\gamma, \sigma, \epsilon)$, respectively, the (partial) Lagrangian of (P0$^\prime\text{-}1$) is given by \eqref{lagrangian}, shown at the top of next page, where $\mv \mu=[\mu_1, \ldots, \mu_L]^T$, $\mv \eta=[\eta_1, \ldots, \eta_L]^T$, $\mv \omega=[\omega_{1,1}, \ldots, \omega_{L,K}]^T$ and $\mv \gamma=[\gamma_{1,1}, \ldots, \gamma_{L,K}]^T$ denote the Lagrangian dual variables associated with the constraints \eqref{times}, \eqref{computing capacity}, \eqref{offloading rate} and \eqref{downloading rate}, respectively. Dual variables $\sigma$ and $\epsilon$ are, respectively, associated with the two constraints specified in \eqref{total upbandwidth} and \eqref{total downbandwidth}. To facilitate primary problem decomposition over $s_l$, \eqref{lagrangian} can be equivalently expressed as \eqref{lagrangian2}.
    
The dual function is thus defined as $g(\mv D)$ as follows
\begin{align}
&~g(\mv D)=\min\limits_{\mv P}\mathcal{L}^{\prime}(\mv P, \mv D)    \label{dual function}  \\
&~\mathtt{Subject \ to}~~\alpha_l^{\rm off}\in[0,1],\alpha_l^{\rm dl}\in[0,1], \forall s_l\in\mathcal{S}.\notag
\end{align}
The corresponding dual problem of $\mathrm{(P0^\prime\text{-}1)}$ is given by 
\begin{subequations}
	\begin{align}
	\mathrm{(D1)}:&~\mathop{\mathtt{Maximize}}~g(\mv D) \notag\\
	&\mathtt {Subject \ to}~~ \mv \mu \geq 0, \mv\eta \geq 0, \mv \omega \geq 0, \label{dual1} \\
	&\mv\gamma \geq 0,
	\sigma \geq 0,
	\epsilon\geq 0.   \label{dual2}
	\end{align}
\end{subequations}

In the following, we solve problem $\mathrm{(P0^\prime\text{-}1)}$ by first evaluating \eqref{dual function} given  fixed $\mv D$, and then iteratively solving problem $\mathrm{(D1)}$ to obtain the optimal solution $\mv D^{\rm opt}$. 

It follows from $\mathcal{L}^{\prime}(\mv P,\mv D)$ (c.f. \eqref{lagrangian2}) that problem \eqref{dual function} can be decomposed into the following subproblems over \(s_l\in\mathcal{S}\):
\begin{subequations} \label{aa}
	\begin{align} 
	&\min_{t_l^{\rm c} \geq 0}~\beta_0\kappa_0(1-\bar{I_l})\frac{(C_lQ_l)^3}{(t_l^c)^2}(1-\prod_{k=1}^K(1-P_{l,k}))\notag\\
	&~~~~~~~+\mu_lt_l^{\rm c}-\eta_lt_l^{\rm c}, \forall s_l\in\mathcal{S};\label{tc}\\
	&\begin{cases}
	\min\limits_{t_{l,\pi(k)}^{\rm dl} \geq 0}~\beta_0p_l^{\rm dl}t_{l,\pi(k)}^{\rm dl}P_{l,\pi(k)}^{\rm dl}+\gamma_{l,k}\frac{R_l}{t_{l,\pi(k)}^{\rm dl}},\\
	~~~~~~~~~~~\forall s_l\in\mathcal{S},k\in\mathcal{K}\backslash\{K\},\\
	\min\limits_{t_{l,\pi(k)}^{\rm dl} \geq 0}~\beta_0p_l^{\rm dl}t_{l,\pi(k)}^{\rm dl}P_{l,\pi(k)}^{\rm dl}+\mu_lt_{l,\pi(k)}^{\rm dl}\\~~~~~~~~~~+\gamma_{l,k}\frac{R_l}{t_{l,\pi(k)}^{\rm dl}},
	 \forall s_l\in\mathcal{S},k=K;
	\end{cases} \;   \\
	&\begin{cases}
	\min\limits_{t_{l,k}^{\rm off} \geq 0}~\beta_kp_k^{\rm off}t_{l,k}^{\rm off}P_{l,k}^{\rm off}+\omega_{l,k}\frac{Q_l}{t_{l,k}^{\rm off}},\\
	~~~~~~~~~\forall s_l\in\mathcal{S},k\in\mathcal{K}\backslash \{K\},\\
	\min\limits_{t_{l,k}^{\rm off} \geq 0}~\beta_kp_k^{\rm off}t_{l,k}^{\rm off}P_{l,k}^{\rm off}+\mu_lt_{l,k}^{\rm off}+\omega_{l,k}\frac{Q_l}{t_{l,k}^{\rm off}},\\
	 ~~~~~~~~~\forall s_l\in\mathcal{S},k=K;
	\end{cases} \;   \label{toff}\\
	&\min_{\alpha_l^{\rm off}\in[0,1]}~\sigma\alpha_l^{\rm off}-\sum_{k=1}^K\omega_{l,k}\alpha_l^{\rm off}B\log_{2}(1+\frac{p_k^{\rm off}u_k}{\alpha_l^{\rm off}}),\notag\\
	&~~~~~~~~~~~~~\forall s_l\in\mathcal{S};  \label{alphaoff}\\
	&\min_{\alpha_l^{\rm dl}\in[0,1]}~\epsilon\alpha_l^{\rm dl}-\sum_{k=1}^K\gamma_{l,k}\alpha_l^{\rm dl}B\log_{2}(1+\frac{p_l^{\rm dl}v_{\pi(k)}}{\alpha_l^{\rm dl}}),\notag\\
	&~~~~~~~~~~~~~\forall s_l\in\mathcal{S}. \label{alphadl}
	\end{align}
\end{subequations}
The optimal solution to subproblem \eqref{tc}-\eqref{toff}, denoted by $(\mv t^{\rm c})^{*}$, $(\mv t^{\rm dl})^*$ and $(\mv t^{\rm off})^*$ is obtained in the following lemma.

\underline{\emph{Lemma}} \emph{4.1}: Given fixed $\mv D$, the optimal solution to \eqref{tc}\text{-}\eqref{toff}, are respectively given by
\begin{subequations}
	\begin{align}
	&(t_l^{\rm c})^*=\begin{cases}
	\left(\tfrac{2\beta_0\kappa_0(1-\bar I_l)(C_lQ_l)^3(1-\prod\limits_{k=1}^K(1-P_{l,k}))}{\mu_l-\eta_l}\right)^\frac{1}{3},\\
	&\!\!\!\!\!\!\!\!\!\!\!\!\!\!\!\!\!\!\!\!\!\!\!\!\!\!\!\!\!\!\!\!\!\!\!\!\!\!\!\!\! \mbox{if} \; \mu_l-\eta_l > 0,\\
	\inf,&\!\!\!\!\!\!\!\!\!\!\!\!\!\!\!\!\!\!\!\!\!\!\!\!\!\!\!\!\!\!\!\!\!\!\!\!\!\!\!\!\! \mbox{otherwise}. \\
	\end{cases} \; \label{tcs}  \\
	&(t_{l,\pi(k)}^{\rm dl})^*=\begin{cases}
	\sqrt{\frac{\gamma_{l,k}R_l}{\beta_0p_l^{\rm dl}P_{l,\pi(k)}^{\rm dl}}},& \forall s_l\in\mathcal{S},k\in\mathcal{K}\backslash \{K\}, \\
	\sqrt{\frac{\gamma_{l,k}R_l}{\beta_0p_l^{\rm dl}P_{l,\pi(k)}^{\rm dl}+\mu_l}},& \forall s_l\in\mathcal{S},k=K. \\
	\end{cases} \; \label{tdls}  \\
	&(t_{l,k}^{\rm off})^*=\begin{cases}
	\sqrt{\frac{\omega_{l,k}Q_l}{\beta_kp_k^{\rm off}P_{l,k}^{\rm off}}},& \forall  s_l\in\mathcal{S},k\in\mathcal{K}\backslash \{K\}, \\
	\sqrt{\frac{\omega_{l,k}Q_l}{\beta_kp_k^{\rm off}P_{l,k}^{\rm off}+\mu_l}}, & \forall s_l\in\mathcal{S},k=K. \\
	\end{cases} \; \label{toffs}  
	\end{align}
\end{subequations}

\emph{Proof:} Please refer to Appendix \uppercase\expandafter{\romannumeral1}.\\

To solve \eqref{alphaoff}, we first take the derivative of its objective function w.r.t. \(\alpha_l^{\rm off}\), denoted by \(F(\alpha_l^{\rm off})\), \(s_l\in\mathcal{S}\), which is defined as follows:
\begin{multline}
F(\alpha_l^{\rm off})=\\
\sum_{k=1}^K\frac{\omega_{l,k}B}{\ln2}\bigg(\ln\big(1+\frac{p_k^{\rm off}u_k}{\alpha_l^{\rm off}}\big)-\frac{p_k^{\rm off}u_k}{\alpha_l^{\rm off}+p_k^{\rm off}u_k}\bigg)-\sigma. 
\end{multline}
It is verified that $F(\alpha_l^{\rm off})$ is non-increasing w.r.t $\alpha_l^{\rm off}\in(0,1]$ with \(\lim\limits_{\alpha_l^{\rm off}\to 0^{+}}F(\alpha_l^{\rm off})=+\infty>0\) and \(F(1)=\sum_{k=1}^K\frac{\omega_{l,k}B}{\ln2}\bigg(\ln\big(1+p_k^{\rm off}u_k\big)-\frac{p_k^{\rm off}u_k}{1+p_k^{\rm off}u_k}\bigg)-\sigma\). Therefore, if $F(1)>0$, it suggests that $F(\alpha_l^{\rm off})>0$ over $\alpha_l^{\rm off}\in(0,1]$, and that the optimal $\alpha_l^{\rm off}$ to \eqref{alphaoff} is \((\alpha_l^{\rm off})^{\ast}=1\); otherwise, there must be some $\tilde\alpha_l^{\rm off}\in(0,1]$ such that $F(\tilde\alpha_l^{\rm off})=0$, which turns out to be the optimal $\alpha_l^{\rm off}$ and can be found numerically via bisection method. To sum up
\begin{align}
(\alpha_l^{\rm off})^{\ast}=\begin{cases}
1, & \mbox{if $F(1)>0$}, \\ 
\tilde\alpha_l^{\rm off}, & \mbox{otherwise}. \label{alphae}
\end{cases}
\end{align}
Applying similar procedure to subproblem \eqref{alphadl}, we can also obtain the optimal $(\alpha_l^{\rm dl})^{\ast}, \forall s_l\in\mathcal{S}$.

Next, we begin solving problem $\mathrm{(D1)}$. Since \eqref{tcs} implies that the optimal dual variables satisfy \(\mu_l-\eta_l> 0\), \(\forall s_l\in\mathcal{S}\), problem $\mathrm{(D1)}$ is recast as below:
\begin{subequations}
	\begin{align}
	\mathrm{(D1{}^\prime)}:&~\mathop{\mathtt{Maximize}}~g(\mv D) \notag\\
	&\mathtt {Subject \ to}~~ \eqref{dual1}, \eqref{dual2}, \notag\\
	&\mu_l-\eta_l > 0, \forall s_l\in \mathcal{S}. \notag
	\end{align}
\end{subequations}
As $g(\mv D)$ is convex but non-differentiable, we iteratively solve $\mathrm{(D1{}^\prime)}$ by subgradient based methods, e.g., (constrained) ellipsoid method, the algorithm of which is summarized in Algorithm  \ref{Algorithm 1} \cite{convex}.
\begin{algorithm}[htp]
	\caption{Ellipsoid Method for Problem (D1${}^\prime$)}\label{Algorithm 1}
	\SetKwInOut{Input}{Input}
	\SetKwInOut{Output}{Output}
	\Input{Dual variables $\mv D^{(0)}$ which is centered at ellipsoid $\mathcal{E}^{(0)} \subset \mathbb{R}^{(2KL+2L+2)\times 1}$ containing the optimal dual solution, $n=0$}
	\Repeat{\emph{the stopping criterion for the ellipsoid method is met}}{
		Obtain $\mv P^*$ based on \eqref{tcs}\text{-}\eqref{toffs} and \eqref{alphae}\;		
		Update the ellipsoid $\mathcal{E}^{(n+1)}$ based on $\mathcal{E}^{(n)}$ and the subgradient of $g(\mv D^{(n)})$ w.r.t. the dual variables \cite{convex}; and set $\mv D^{(n+1)}$ as the center of ellipsoid $\mathcal{E}^{(n+1)}$\;   Set $n=n+1$.	
	}	
	\Output{$\mv D^{\rm opt} \leftarrow$ $\mv D^{(n)}$}
\end{algorithm}

It then remains to find the primal-optimal solution to $\mathrm{(P0^\prime\text{-}1)}$. Since $(t_l^{\rm c})^*$, $(t_{l,\pi(k)}^{\rm dl})^*$, $(t_{l,k}^{\rm off})^*$, $(\alpha_l^{\rm off})^*$ and $(\alpha_l^{\rm dl})^*, \forall k\in\mathcal{K}, s_l\in\mathcal{S}$ are unique optimal solution to problem \eqref{tc} - \eqref{alphadl}, the optimal solution $(t_l^{\rm c})^{\rm opt}$, $(t_{l,\pi(k)}^{\rm dl})^{\rm opt}$, $(t_{l,k}^{\rm off})^{\rm opt}$ to $\mathrm{(P0^\prime\text{-}1)}$ can be directly obtained by plugging $\mv D^{\rm opt}$ into \eqref{tcs} - \eqref{toffs}, while the optimal solutions $(\alpha_l^{\rm off})^{\rm opt}$ and $(\alpha_l^{\rm dl})^{\rm opt}$ are numerically attained (c.f.~\eqref{alphae}). To sum up, with any (feasible) caching decisions given, problem  $\mathrm{(P0^\prime\text{-}1)}$ can be solved by the  dual decomposition method as above.   

The optimal solution to $\mathrm{(P0^\prime)}$ can be found by exhaustive search with high computational complexity of $\mathcal{O}(2^{|\mathcal{S}|})$. To accommodate large number of services $|\mathcal{S}|$ with UEs at different locations having independent request over the service library \(\mathcal{S}\), we propose in general a DL-based algorithm to find cache placement for $\mathrm{(P0^\prime)}$ in the next section.

\section{DL-based Offline Cache Placement} \label{sec:Proposed solution}
The optimal cache placement shall balance various coupled factors of the tasks such as popularity, uplink and/or downlink data transmission quality, data size and computation intensity. As a result, to avoid numerically solving complex MINLP for optimal caching decisions every time the channel or the task information changes, in this section, we propose a DL-based hybrid learning framework to solve $\mathrm{(P0^\prime\text{-}2)}$.

We consider fixed distance between the service locations and the BS with the channel coefficients distributed as $h_k^\prime$ $(g_k^\prime)\sim\mathcal{CN}(0, A_0(d_0/d_k)^\gamma)$ and thus the normalized channel gains $u_k$'s $(v_k$'s$)$ following exponential distribution with parameters \(\frac{N_0B}{A_0(d_0/d_k)^\gamma}\) $\big(\frac{N_0B}{A_0(d_0/d_k)^\gamma}\big)$. We also assume that the input/output bit-length for computation tasks in \(\mathcal{S}\) are drawn from uniform distributions denoted as $\mathcal{U}(a,b)$, where $a$ and $b$ are the minimum and maximum bounds of the distributions, respectively. {\color{black}As a result, a sufficient number of data samples composed of quadruples as $(\mv h^{(t)}, \mv g^{(t)}, \mv Q^{(t)}, \mv R^{(t)})$, where \(\mv h^{(t)}=(u_1,\ldots, u_K)^T\), \(\mv g^{(t)}=(v_1,\ldots, v_K)^T\), \(\mv Q^{(t)}=(Q_1,\ldots, Q_L)^T\), and \(\mv R^{(t)}=(R_1,\ldots, R_L)^T\) can be synthesized offline in the $t$th iteration, while the corresponding caching decisions \( \mv I^{(t)\ast}=(I^{(t)\ast}_1, \ldots, I^{(t)\ast}_L)^T\) serving as ``labels'' are generated during the training as going to be introduced  shortly. When the training finishes, whenever a change arises in the input quadruple, the trained model can be evaluated to yield the joint solution of resource allocation and cache placement.}

\begin{figure*}[htp]
	\centering
	\includegraphics[width=7.1in]{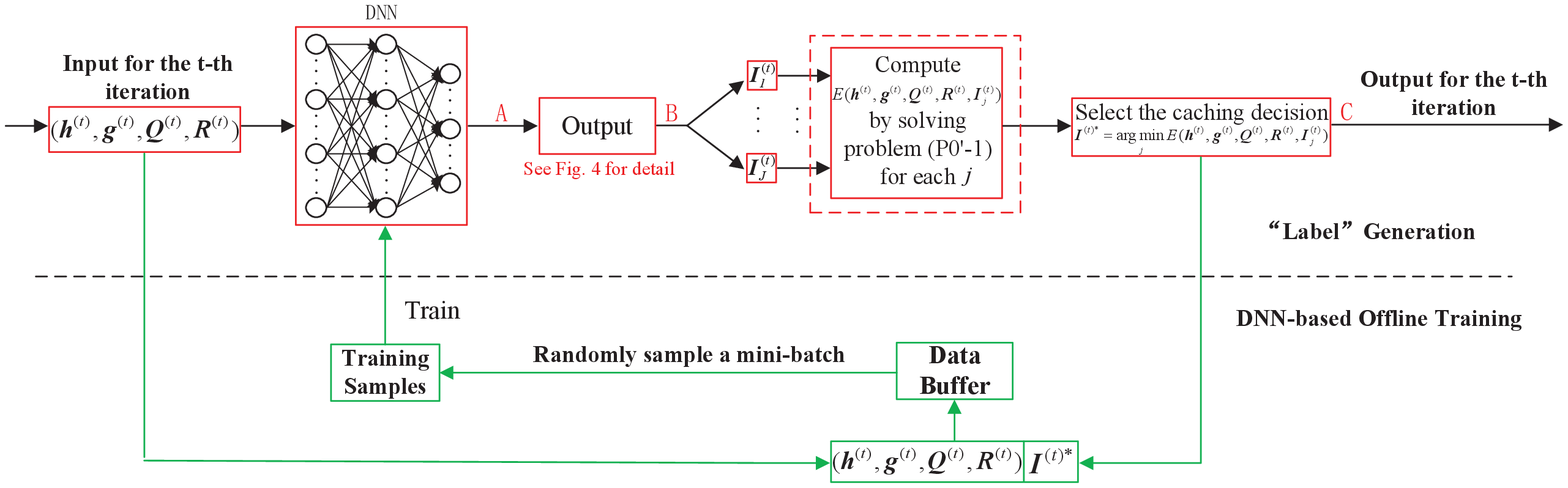}
	\textcolor{black}{\caption{The offline learning framework for joint resource allocation and cache placement.}}
	\label{fig:DNN}
	
	\centering
	\includegraphics[width=6.6in]{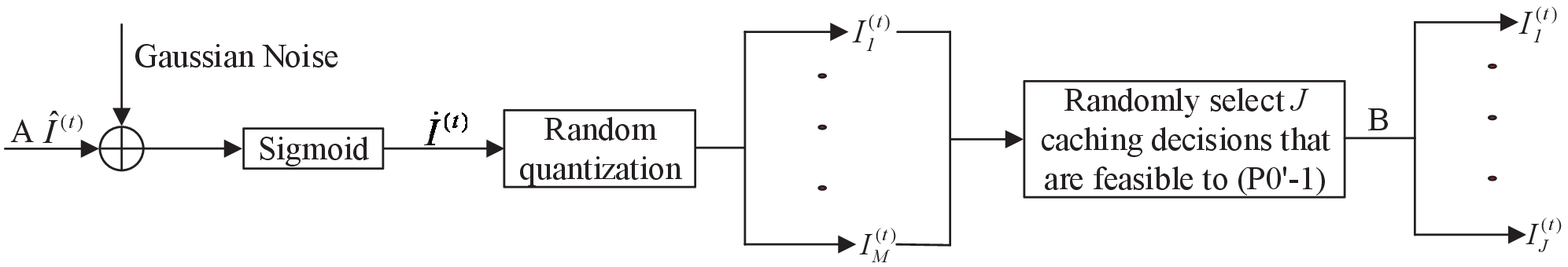}
	\textcolor{black}{\caption{Stochastic quantization method.}}
	\label{fig:q}
\end{figure*}

Mathematically, our goal is to generalize a nonlinear mapping between sample inputs and outputs using an approximation function \(f_{\mv \theta^{(t)}}\) parameterized by \(\mv \theta^{(t)}\) via a DNN, which is defined as: 
\begin{align}
\dot{\mv I}^{(t)}=f_{\mv \theta^{(t)}}(\mv h^{(t)}, \mv g^{(t)}, \mv Q^{(t)}, \mv R^{(t)}).
\end{align}
In order to identify the parameter vector \(\mv\theta^{(t)}\) for the mapping $f_{\mv \theta^{(t)}}(\mv h^{(t)}, \mv g^{(t)}, \mv Q^{(t)}, \mv R^{(t)})$, we formulate a learning problem with the empirical risk function that measures the mean-square error (MSE) between the model output \(f_{\mv \theta^{(t)}}\) and the (labelled) caching decisions $\mv I^{(\omega)\ast}$ as the objective {\color{black}(referred as ``training loss'', evaluated at ``C'' in Fig. 3):}
\begin{subequations}
	\begin{align}
	&~\mathrm{(P1)}:\notag\\
	&~\mathop{\mathtt{Minimize}}_{\mv \theta}\!\!\!
	~~\mb{E}_{\omega}\lVert \mv I^{(w)\ast}-f_{\mv \theta^{(t)}}(\mv h^{(\omega)}, \mv g^{(\omega)}, \mv Q^{(\omega)}, \mv R^{(\omega)})\lVert^2,\notag
	\end{align}
\end{subequations} 
where \(\omega\) denotes the index of the training sample.
The offline learning framework for solving problem $\mathrm{(P0^\prime\text{-}2)}$ is summarized in Fig.~\ref{fig:DNN}. It consists of two alternating stages: service caching decisions (``labels") generation and DNN based offline training, which are detailed in the following subsections.

\subsection{Service Caching Decisions Generation}
To find the optimal cache placement for problem $\mathrm{(P0^\prime\text{-}2)}$ by solving $\mathrm{(P0^\prime\text{-}1)}$ requires exhaustive search over \(2^L\) binary candidates, thus causing complexity of \(\mathcal{O} \big(2^L\times 2(2KL+2L+2)^2 \log (\sqrt{\phi}W/\zeta) \big)\), where \(2(2KL+2L+2)^2\) accounts for the complexity for solving $\mathrm{(P0^\prime\text{-}1)}$ using Algorithm \ref{Algorithm 1} \cite{ellipsoid}, $W\triangleq \max\limits_{\mv \omega\in \partial g(\mv D), \mv D\in \mathcal{E}^{(0)}}\Vert{\mv \omega}\Vert$ is a Lipschitz constant for \eqref{dual function} over the initial ellipsoid $\mathcal{E}^{0}=\{\mv D|\Vert \mv D \Vert \leq \sqrt{\gamma}\}$, $\mv \omega$ is a sub-gradient of $g(\mv D)$ over $\mathcal{E}^{(0)}$, and $\zeta$ is a parameter controlling the accuracy of the ellipsoid algorithm. To address this challenge, we propose in this subsection a suboptimal ``label" generation scheme that aims for ``exploitation'' of the current DNN outputs $\dot{\mv I}^{(t)}$ while providing sufficient diversity for ``exploration''. To generate feasible service cache placement, we quantize $\dot{\mv I}^{(t)}$ into a number $J$ of candidates. Specifically, we propose a stochastic quantization mapping defined as 
\begin{align}
g_J:\dot{\mv I}^{(t)} \mapsto \{\mv I_{j}^{(t)}|\mv I_{j}^{(t)}\in\{0,1\}^L,j=1,\ldots,J\},
\end{align}
which is illustrated in Fig. \ref{fig:q}.

To elaborate, first, we add Gaussian noise to the DNN logits $\hat{\mv I}^{(t)}$ to generate more diversity in the caching decision space \(\{0, 1\}^L\). The activation operating element-wise on the noisy logits can be expressed as $\dot{\mv I}^{(t)}=f_{sg}(\hat{\mv I}^{(t)}+\mv n)$, where $\mv n\sim C\mathcal{N}(0,\mv I)$, {\color{black}where $f_{sg}(\cdot)$ is the $\mathrm{sigmoid}$ function defined by $f_{\rm sg}(x)=1/(1+e^{-x})$, such that $\dot{I}_l^{(t)}$, the $l$th entry of $\dot{\mv I}^{(t)}$, falls with in $[0,1]$.} Next, we sample from Bernoulli distribution a binary caching decision for each of the $L$ services $M$ times: 
\begin{align}
I_{l,m}^{(t)}=\begin{cases}
1, & \mbox{with probability $\dot{I}^{(t)}_l$}, \\ 
0, & \mbox{otherwise}, \label{sampling}
\end{cases}
\end{align}
where $I_{l,m}^{(t)}$ denotes the $l$th entry of the $m$th candidate \(\mv I_m^{(t)}=(I_{1,m}^{(t)},\ldots, I_{L,m}^{(t)})^T\), $\forall m\in\{1,\ldots, M\}$. Finally, we randomly select $J$ $(J < M)$ out of $M$ caching decisions that satisfy constraint \eqref{C:caching capacity constraint}.\footnote{If there are not enough feasible caching decisions, we randomly select a missing number of candidates from the rest of $M$ caching decisions or resample $M$ sets of caching decisions until $J$ candidate caching decisions are generated.} \textcolor{black}{To take into account candidate solutions with different noise-weighting, we include in every selected $J$ sets of candidate caching decisions one candidate where each entry $I_l^{(t)}$ is sampled from $\dot{I_l}^{(t)}=f_{sg}(\hat{I_l}^{(t)})$ based on noise-free logits, \(l=1,\ldots, L\).} Then we evaluate the performance of the $J$ candidate decisions by optimally solving $\mathrm{(P0^\prime\text{-}1)}$ (see Section \ref{sec:Optimal Solution} for detail) using off-the-shelf convex problem solvers such as CVX \cite{cvx}, and the one with the minimum expected energy consumption $E(\mv h^{(t)}, \mv g^{(t)}, \mv Q^{(t)}, \mv R^{(t)}, \mv I^{(t)}_j)$ will be selected as the caching decision \(\mv I^{(t)\ast}\) serving as ``labels" for problem \(\mathrm{(P1)}\). 
\begin{remark} 
	For the determined order-preserving based caching decisions generation employed in \cite{liang}, the orders for any two entries are preserved across all $M$ candidates. That is, if \(\hat{I}_{l_1}^{(t)}\le\hat{I}_{l_2}^{(t)}\), then \( I_{l_1,m}^{(t)}\le I_{l_2,m}^{(t)}\) for any \(m\in\{1,\ldots M\}\). By comparison, the sampled caching decisions based on \eqref{sampling} provide more uncertainties, and therefore, by properly choosing $M$ and $J$, it is more likely to find promising candidates satisfying the constraints \eqref{C:caching capacity constraint}.
\end{remark}

\begin{remark}
\textcolor{black}{Note that the choice of $J$ plays an important role in balancing the quality of the output caching decision and computational complexity. Since the selected (feasible) caching decisions are evaluated to approximate the ground-truth label (C in Fig. 3), given a sufficiently large $M$ fixed, larger $J$ implies more reliable approximation and thus faster training convergence at the cost of higher per-iteration computation complexity, and vice versa.}
\end{remark}

Note that any candidate caching decisions violating constraint \eqref{C:caching capacity constraint} are discarded. By examining problem $\mathrm{(P0^\prime)}$, the caching decisions achieve optimality when constraint \eqref{C:caching capacity constraint} is satisfied to its maximum extent.\footnote{The constraint \eqref{C:caching capacity constraint} being satisfied to its maximum extent refers to caching decisions that are feasible to problem $\mathrm{(P0^\prime)}$, but incurs violation of \eqref{C:caching capacity constraint} if any more type of service is cached.} Inspired by this, we sort the entries of all eligible candidate solutions in descending orders according to their corresponding value in $\dot{\mv I}^{(t)}$, and flip $0$'s to $1$'s until \eqref{C:caching capacity constraint} is satisfied to its maximum extent. 
 
\subsection{DL-Based Offline Training}
The newly obtained ``label" \(\mv I^{(t)\ast}\), combined with the sample-inputs \((\mv h^{(t)}, \mv g^{(t)}, \mv Q^{(t)}, \mv {R}^{(t)})\), forms a new input-``label" pair \(( \mv h^{(t)}, \mv g^{(t)}, \mv Q^{(t)}, \mv {R}^{(t)}, \mv I^{(t)\ast} )\). Specifically, we start with training after sufficient number of input-``label" samples are collected in the data buffer, and update the parameter vector \(\mv\theta^{(t)}\) every \(\tau\) iteration by a stochastic gradient descent (SGD) step as follows
\begin{align}
\mv\theta^{(t+1)}=\mv\theta^{(t)}-\eta^{(t)}\hat{\nabla}L(\mv\theta^{(t)}), \label{update}
\end{align}
where \(\eta^{(t)}\) is the learning rate, and \(\hat{\nabla}L(\mv\theta^{(t)})=\tfrac{1}{|\mathcal{D}^{(t)}|}\sum_{\omega\in\mathcal{D}^{(t)}}\nabla L(\mv\theta^{(t)};\mv h^{(\omega)}, \mv g^{(\omega)}, \mv Q^{(\omega)}, \mv R^{(\omega)})\) is the stochastic gradient approximating \(\mb{E}_{\omega}[\nabla L(\mv\theta^{(t)};\mv h^{(\omega)}, \mv g^{(\omega)}, \mv Q^{(\omega)}, \mv R^{(\omega)})]\) via a mini-batch \(\mathcal{D}^{(t)}\) of samples from the data buffer in the $t$th iteration. Furthermore, we also maintain the data buffer with limited capacity, where only the latest $|\mathcal{R}|$ input-``label" pairs are kept for model updates. The overall DL-based cache placement algorithm is summarized in Algorithm \ref{table:Algorithm 2}. 

\begin{algorithm}[htp]
\caption{DL-Based Offline Cache Placement}\label{table:Algorithm 2}	
\SetKwInOut{Input}{Input}	
\SetKwInOut{Output}{Output}
Initialize the parameter vector $\mv\theta^{(0)}$\; 
\Repeat{\emph{the algorithm converges}}{	
	\Input{Wireless channel gains $\mv h^{(t)}$ and $\mv g^{(t)}$, service input-bit size $\mv Q^{(t)}$ and output-bit size $\mv R^{(t)}$ at each iteration $t$}	
	Obtain the model output $\dot{\mv I}^{(t)}=f_{\mv \theta^{(t)}}(\mv h^{(t)}, \mv g^{(t)}, \mv Q^{(t)}, \mv R^{(t)})$\;		
	Generate a set of $J$ feasible caching decisions $\mv I_j^{(t)}=g_J(\dot{\mv I}^{(t)}), \forall j\in \{1,\ldots,J\}$\;		
	Compute $E(\mv h^{(t)}, \mv g^{(t)}, \mv Q^{(t)}, \mv R^{(t)},\mv I_{j}^{(t)})$ for all $\mv I_{j}^{(t)}$ by solving $\mathrm{(P0^\prime\text{-}1)}$\; 
	Select the best caching decision $\mv I^{(t)\ast}=\arg \min\limits_jE(\mv h^{(t)}, \mv g^{(t)}, \mv Q^{(t)}, \mv R^{(t)},\mv I^{(t)}_j)$\;
	Feed the data buffer using the newly collected input-``label" pair $(\mv h^{(t)}, \mv g^{(t)}, \mv Q^{(t)}, \mv R^{(t)},\mv I^{(t)\ast})$\;		
	\If{$t$ mod $\tau=0$}{
		Randomly sample a mini-batch \(\mathcal{D}^{(t)}\) of samples $\{(\mv h^{(\omega)},\mv g^{(\omega)}, \mv Q^{(\omega)}, \mv R^{(\omega)},\mv I^{(\omega)\ast})|\omega\in\mathcal{D}^{(t)}\}$ from the data buffer\;
		{\color{black}Update $\mv \theta^{(t)}$ using SGD step (c.f.~\eqref{update}) with momentum optimizer \cite{2013sut}};
	}		
}	
\end{algorithm}

After the training converges, given any input \((\mv h^{(t)}, \mv g^{(t)}, \mv Q^{(t)}, \mv {R}^{(t)})\), the caching decisions can be obtained by implementing steps 3-6 in Algorithm \ref{table:Algorithm 2}, whose model-inference  complexity is \(\mathcal{O} \big(J\times 2(2KL+2L+2)^2 \log (\sqrt{\phi}W/\zeta) \big)\).
\begin{remark}
Compared to a fully model-free DL approach which incorporates all binary caching decision and continuous resource allocations as actions, the proposed hybrid learning framework significantly reduces the action space by simplifying the learning task to a classification problem. Moreover, the model-based optimization module facilitates faster convergence than a fully model-free DL approach, whose training is often compromised by inaccurate evaluation of the actions due to insufficient training. In addition, after model deployment, the proposed hybrid approach obtains effective cache placement and the corresponding resource allocation by model inference with little computation overhead.
\end{remark}
\begin{remark}
\textcolor{black}{While the proposed learning framework can be used offline by synthesizing sufficient amount of input samples based on full knowledge of the distribution of channel gains and task-input/output data size, it can also serve as an online training framework to accommodate application scenarios when input samples can only be revealed real-time with partial or no knowledge of their (stable) distributions.}
\end{remark}

\section{Special Case} \label{special}
Consider special scenarios in which UEs at each location demand a unique type of service in $\mathcal{S}^\prime=\{s_1,\ldots, s_K\}\subseteq\mathcal{S}$. This is equivalent to \(P_{l,k}=1\) for \(s_l=s_k\) and \(P_{l,k}=0\) for \(s_l\in\mathcal{S^\prime}\setminus\{s_k\}\), \(\forall k\in\mathcal{K}\). Indices \(l\) and \(k\) thus become interchangeable. Hence, $E^{\rm c}$ is recast as $\sum_{k\in\mathcal{K}}\kappa_0\Myfrac{\left(C_kQ_k\right)^3(1-I_k)}{\left(t_k^{\rm c}\right)^2}$ (c.f. \eqref{eq:BS computation energy}); $E_k^{\rm off}$ reduces to $p_k^{\rm off}t_k^{\rm off}(1-I_k)$ (c.f. \eqref{eoff}); and $E^{\rm dl}$ is simplified as $\sum_{k\in\mathcal{K}}t_k^{\rm dl}p_k^{\rm dl}$ (c.f. \eqref{downloading energy}). In addition, constraints \eqref{time} also reduce to $t^{\rm off}_k+t_k^{\rm c}+t_{k}^{\rm dl} \le T_k, k\in\mathcal{K}$. 
By denoting \(\mv I=[I_1, \ldots, I_K]^T\), \(\mv \alpha^{\rm off}=[\alpha_1^{\rm off}, \ldots, \alpha_K^{\rm off}]^T\), \(\mv \alpha^{\rm dl}=[\alpha_1^{\rm dl}, \ldots, \alpha_K^{\rm dl}]^T\), \(\mv t^{\rm c}=[t_1^{\rm c}, \ldots, t_K^{\rm c}]^T\), \(\mv t^{\rm off}=[t_1^{\rm off}, \ldots, t_K^{\rm off}]^T\) and $\mv t^{\rm dl}=[t_1^{\rm dl}, \ldots, t_K^{\rm dl}]^T$, the weighted-sum energy minimization problem under this special circumstance is formulated as \cite{my}
\begin{subequations}
	\begin{align}
	\mathrm{(P2)}:\!\!&~\mathop{\mathtt{Minimize}}_{\mv I,\mv \alpha^{\rm off},\mv \alpha^{\rm dl} \atop \mv t^{\rm c}, \mv t^{\rm dl}, \mv t^{\rm off}}\!\!\!
	~~\beta_0\sum\limits_{k\in\mathcal{K}}\left(\kappa_0\frac{\left(C_kQ_k\right)^3(1-I_k)}{\left(t_k^{\rm c}\right)^2}+t_k^{\rm dl}p_k^{\rm dl}\right)\notag\\
	&~~~~~~~~~~~~~~~+\sum\limits_{k\in\mathcal{K}}\beta_k p_k^{\rm off}t_k^{\rm off}(1-I_k)\notag\\
	&~~\mathtt{Subject \ to}~~ \sum_{k=1}^KI_kR_k \leq S,\\
	&~~t^{\rm off}_k+t_k^{\rm c}+t_{k}^{\rm dl} \le T_k, \; \forall k\in\mathcal{K}, \label{s dealine}\\
	&~~t_k^{\rm c} \geq \frac{C_kQ_k(1-I_k)}{f_0^{\max}},\forall k\in\mathcal{K}, \label{s computing capacity}\\
	&~~\sum_{k\in\mathcal{K}} \alpha_k^{\rm off} \leq 1,\label{total bandwidth} \\ &~~\sum_{k\in\mathcal{K}}\alpha_k^{\rm dl} \leq 1, \label{total bandwidth dl} \\
	&~~\alpha_k^{\rm{off}} B\log_{2}\left(1+{\frac{p_k^{\rm off} u_k}{\alpha_k^{\rm{off}}}}\right) \geq \frac{Q_k(1-I_k)}{t_k^{\rm off}},  \forall k\in\mathcal{K}, \label{s offloading rate}  \\
	&~~\alpha_k^{\rm{dl}} B\log_{2}\left(1+{\frac{p_k^{\rm{dl}} v_k}{\alpha_k^{\rm{dl}}}}\right) \geq \frac{R_k}{t_k^{\rm dl}}, \forall k\in\mathcal{K}, \label{s downloading rate} \\
	&~~I_k=\left\{0,1\right\}, \alpha_k^{\rm off}\in[0,1], \alpha_k^{\rm dl}\in[0,1], \forall k\in\mathcal{K}.\label{s caching decision}
	\end{align}
\end{subequations}
\begin{remark}\textcolor{black}{
We provide the model-based solution to problem $\mathrm{(P2)}$ for the following reasons. First, given the dual variables, the optimal resource allocation in special case admits (semi-) closed-form solution (c.f.~\eqref{alphaoffs}-\eqref{alphadls}) compared to the general case when BW solution can only be numerically attained. This helps obtain insights for optimal BW allocation. Second, the special case allows fast acquisition of (suboptimal) resource allocation without iterative primal-dual updates thanks to the structure of the semi-closed form solution. In addition, the special case warrants low-complexity solution to the MINLP leveraging ILP without going through the DRL training process. The effect of this solution will be corroborated by numerical results in Section VII. B.}
\end{remark}

We then provide analytical solution in semi-closed form for the special-case problem $\mathrm{(P2)}$ and draw some insights therein. With the caching decisions fixed as $\mv I=\bar{\mv I}$, problem $\mathrm{(P2)}$ reduces to $\mathrm{(P2\text{-}1)}$, whose Karush-Kunh-Tucker (KKT) solution is obtained leveraging the following lemma.

\underline{\emph{Lemma}} \emph{6.1}: By denoting the Lagrangian multiplier associated with constraints \eqref{s dealine}, \eqref{s computing capacity}, \eqref{s offloading rate}, \eqref{s downloading rate}, \eqref{total bandwidth} and \eqref{total bandwidth dl}, by \(\mv \mu=[\mu_1,\cdots,\mu_K]^T\), \(\mv \eta=[\eta_1,\cdots,\eta_K]^T\), \(\mv \omega=[\omega_1,\cdots,\omega_K]^T\), \(\mv \gamma=[\gamma_1,\cdots,\gamma_K]^T\), $\sigma$ and \(\epsilon\), respectively, the KKT solution to problem $\mathrm{(P2\text{-}1)}$ for given \((\mv \mu, \mv \eta, \mv \omega, \mv \gamma, \sigma, \epsilon)\) is as follows:
\begin{subequations}
	\begin{align}
	&\left(t_k^{\rm dl}\right)^*=\sqrt{\frac{\gamma_kR_k}{\beta_0p_k^{\rm dl}+\mu_k}},  \label{tkdl} \\
	&\left(t_k^{\rm off}\right)^*=\sqrt{\frac{\omega_kQ_k}{\beta_kp_k^{\rm off}+\mu_k}(1-\bar I_k)},  \label{toffss}\\
	&(t_k^{\rm c})^*=\begin{cases}
	\left(\frac{2\kappa_0\beta_0(1-\bar I_k)(C_kQ_k)^3}{\mu_k-\eta_k}\right)^\frac{1}{3},& \mbox{if} \; \mu_k-\eta_k > 0,\\
	\inf,& \mbox{otherwise},
	\end{cases} \;   \\
	&(\alpha_k^{\rm off})^*=\min\left\{\frac{p_k^{\rm off}u_k}{e^{\left[W_0\left(-e^{\phi_k \ln 2}\right)-\phi_k \ln 2\right] }-1} , 1 \right\},   \label{alphaoffs} \\
	&(\alpha_k^{\rm dl})^*=\min\left\{\frac{p_k^{\rm dl}v_k}{e^{\left[W_0\left(-e^{\varphi_k \ln 2}\right)-\varphi_k \ln 2\right] }-1} , 1 \right\},  \label{alphadls}
	\end{align}
\end{subequations}
where $W_0\left(\cdot \right)$ is the principal branch of Lambert W function defined as the inverse function of $xe^x=y$ \cite{1996:W}, $\phi_k=-\frac{\sigma}{\omega_k B}-\frac{1}{\ln2}$, and $\varphi_k=-\frac{\epsilon}{\gamma_kB}-\frac{1}{\ln2}$.  

\emph{Proof:} Please refer to \cite[Appendix 1]{my}.
\begin{remark}
	Compared with KKT solution to problem $\mathrm{(P0^\prime\text{-}1)}$ (cf. \eqref{tcs} - \eqref{toffs} and \eqref{alphae}), the optimal offloading/downloading BW for given dual variables can be obtained in semi-closed forms, from which we have the following observations. 1) With the transmitting power $p_k^{\rm off}$ of UEs at location $k$ fixed, the \((\alpha_k^{\rm off})^\ast\) of BW allocated for these UEs for offloading is proportional to their channel gain $h_k$ to the BS, and when $h_k$ increases to be larger than a threshold $\Myfrac{(e^{\left[W_0\left(-e^{\phi_k \ln 2}\right)-\phi_k \ln 2\right] }-1)}{p_k^{\rm off}}$, UEs at location $k$ will gain access to full BW to save transmission latency and thus energy. 2) Likewise, \((\alpha_k^{\rm off})^\ast\) is also increasing with $p_k^{\rm off}$ such that the UEs with larger transmitting power are able to finish task offloading faster to save energy. Similar insights can also be drawn from \eqref{alphadls}. 
\end{remark}

Note that for the special-case problem $\mathrm{(P2)}$ only, we propose an ILP-based suboptimal cache placement scheme leveraging the KKT solution for BW allocation given by \eqref{alphaoffs} and \eqref{alphadls}. First, under the assumption that the computation frequency of the edge server is fully used for each computation task, e.g., $f_k = f_0^{\rm max}$, $\forall k\in\mathcal{K}$, the execution delay of any task is highly probably shorter than the deadline for the purpose of energy saving, i.e., $t_k^{\rm off}+t_k^c+t_k^{\rm dl}< T$, $\forall k\in \mathcal{K}$. The optimal dual variables associated with constraints \eqref{s dealine} thus become zero due to the complementary slackness. Then assuming that there is no cache placed for any tasks, i.e., \(I_k=0\), $\forall k\in\mathcal{K}$, we substitute \eqref{toffss} and \eqref{alphaoffs} for \(t_k^{\rm off}\) and \(\alpha_k^{\rm off}\), respectively, in \eqref{total bandwidth} and \eqref{s offloading rate}. Since it is easy to verify that \eqref{total bandwidth} and \eqref{s offloading rate} are achieved active for optimal solution to $\mathrm{(P2\text{-}1)}$, this implies a set of equations as follows.
\begin{align}
\begin{cases}
f(\omega_k,\sigma)=0, \forall k\in\mathcal{K},\\ 
g(\omega_1,\cdots,\omega_K,\sigma)=0, \label{eq:set of equations}
\end{cases}
\end{align}
where $\forall k\in\mathcal{K}$,
\begin{small}
\begin{align}
&f(\omega_k,\sigma)=\frac{Bp_k^{\rm off}u_k}{\ln 2 \sqrt{O_k\beta_kp_k^{\rm off}}}\notag\\
&-\frac{\exp\Big(W_0\big(-\exp(\phi_k(\omega_k, \sigma)\ln 2)\big)-\phi_k(\omega_k, \sigma) \ln 2\Big)-1}{\Big(W_0\big(-\exp(\phi_k(\omega_k, \sigma) \ln 2)\big)-\phi_k(\omega_k, \sigma) \ln 2\Big)\sqrt{\omega_k}},\notag\\ 
&g(\omega_1,\cdots,\omega_K,\sigma)=\notag\\
&\sum\limits_{k\in\mathcal{K}}\frac{p_k^{\rm off}u_k}{\exp\Big(W_0\big(-\exp(\phi_k(\omega_k, \sigma) \ln 2)\big)-\phi_k(\omega_k, \sigma) \ln 2\Big)-1}\notag\\&-1. \notag
\end{align}
\end{small}
\underline{\emph{Lemma}} \emph{6.2}: There must exist numerical solutions of $\sigma$ and $\omega_k$, $\forall k\in\mathcal{K}$  to the set of equations in \eqref{eq:set of equations}. 

\emph{Proof:} $f(\omega_k,\sigma)$ and $g(\omega_1,\cdots,\omega_K,\sigma)$ are both non-decreasing w.r.t $\omega_k$ and non-increasing w.r.t $\sigma$, $\forall k\in\mathcal{K}$ (Please refer to \cite[Appendix 2]{my}). Moreover, it is easily verified that \(\lim\limits_{\omega_k\to 0^{+}}f(\omega_k,\sigma)=-\infty<0\), $k\in\mathcal{K}$, and \(\lim\limits_{\sigma\to 0^{+}}g(\omega_1,\cdots,\omega_K,\sigma)=+\infty>0\). Based on the monotonicity of the two functions, we use bi-section method to solve $f(\omega_k, \sigma)=0$ by fixing $\sigma$, and then plug the solution $\omega_k$, $\forall k\in \mathcal{K}$, into $g(\omega_1,\cdots,\omega_K,\sigma)$ to further find $\sigma$ via bi-section until $g(\omega_1,\cdots,\omega_K,\sigma) = 0$ is met. 

We can solve another similar set of equations as \eqref{eq:set of equations} to obtain optimal $\epsilon$ and $\gamma_k$, $\forall k\in\mathcal{K}$. Then with \((\alpha_k^{\rm off})^{\ast}\)'s and \((\alpha_k^{\rm dl})^{\ast}\)'s numerically obtained (c.f.~\eqref{alphaoffs} and \eqref{alphadls}), $t_k^{\rm c}=\Myfrac{C_kQ_k}{f_0^{\max}}$, \(t_k^{\rm off}\) (c.f.~\eqref{s offloading rate}) and \(t_k^{\rm dl}\) (c.f.~\eqref{s downloading rate}), \(k\in\mathcal{K}\), are obtained as constants, denoted by \(\bar t_k^{\rm c}\), \(\bar t_k^{\rm off}\) and \(\bar t_k^{\rm dl}\), respectively. As a result, problem $\mathrm{(P2)}$ reduces to an ILP, with only the caching decision \(\mv I\) as optimization variables as follows:
	\begin{align}
	\mathrm{(P2\text{-}2)}:\!\!&~\mathop{\mathtt{Minimize}}_{\mv I}\!\!\!
	~~\beta_0\sum\limits_{k\in\mathcal{K}}\kappa_0\frac{\left(C_kQ_k\right)^3(1-I_k)}{\left(\bar t_k^{\rm c}\right)^2} \notag\\
	&~~~~~~~~~~~~~+\sum\limits_{k\in\mathcal{K}}\beta_k p_k^{\rm off}\bar t_k^{\rm off}(1-I_k)\notag\\
	&~~\mathtt{Subject \ to}~~ \sum_{k=1}^KI_kR_k \leq S, ~I_k=\left\{0,1\right\},  \forall k\in\mathcal{K}. \notag
	\end{align}
\begin{remark}
	The ILP problem $\mathrm{(P2\text{-}2)}$, despite of being exponentially complex in the worst case, admits complexity of \(\mathcal{O}(L^2\log L)\) on average thanks to the recently developed fast branch and bound method, e.g., Lenstra-Lenstra-Lovasz (LLL) algorithm \cite{2009:m}, which can be effectively solved using off-the-shelf software packages, e.g., \cite{gurobi}. In addition, the objective function of problem $\mathrm{(P2\text{-}2)}$ suggests to cache the results of  those tasks requiring high energy consumption in task offloading and computation.
\end{remark}

When the ILP-based cache placement is obtained by solving $\mathrm{(P2\text{-}2)}$, we solve $\mathrm{(P2\text{-}1)}$ once again to get the corresponding suboptimal resource allocation. 

\section{Numerical Results} \label{sec:Numerical Results}
In this section, we verify the effectiveness of our proposed DL-based service cache placement algorithms for problem $\mathrm{(P0)}$ as well as the suboptimal cache placement designed for the special-case problem $\mathrm{(P1)}$. 
We consider a wireless setup where there are $K=5$ locations deployed on a circle with radius $d_k=d=0.03$ km centered on the BS, \(\forall k\in\mathcal{K}\), and a service library with $L=10$ types of services. A task request from location \(k\in\mathcal{K}\) is assumed to follow \emph{Zipf} distribution given by {\color{black}\cite{2016:S} -\cite{2019yan}}
\begin{align}
P_{l,k}=\frac{1}{l_k^{\sigma_k}}\left(\sum\limits_{s_l\in\mathcal{S}}\frac{1}{l^{\sigma_k}}\right)^{-1}, \label{plk}
\end{align}
where $\sigma_k=0.9$ determines the skewness of the preference profile at location $k$, and \(l_k=\pi_k(l)\) is the rank of service \(s_l\in\mathcal{S}\) in terms of popularity at location $k$, represented by a permutation $\pi_k(\cdot)$ over $\mathcal{S}$. The average channel gain $A_0$ is set as $-128.1$ dB at reference distance $d_0=1$ km with the pathloss exponent factor $\gamma=2.6$ \cite{2019:bi}. The Rayleigh fading is generated by $i.i.d.$ complex Gaussian RVs with zero mean and unit variance. The task-input and task-output bit-lengths follow uniform distributions, denoted by $Q_l\sim \mathcal{U}[7,7.5]$ Mbits and $R_l\sim \mathcal{U}[21,22]$ Mbits, $s_l \in \mathcal{S}$. Other parameters are set as follows unless otherwise specified: transmission BW $B_{\rm off}=B_{\rm dl}=10$ MHz, noise spectrum density \(N_0=-169\) dBm/Hz, weight factors for problem $\mathrm{(P0)}$ $\beta_0=0.5$, $\beta_k=0.5/K$, the maximum edge server's computation frequency $f_0^{\rm max}=10$ GHz, transmission power $p_k^{\rm off}=0.25$ W, $p_l^{\rm dl}=1$ W, capacitance coefficient $\kappa=10^{-27}$, and the number of CPU cycles required for computation service $s_l$ $C_l=1000$ cycles/bit, $\forall k\in\mathcal{K}, s_l\in\mathcal{S}$ \cite{2019:h}. Furthermore, the deadline for each task is set to be the same, e.g., $T_l=T=2.8$ s, $\forall s_l\in\mathcal{S}$, and the caching capacity $\mathcal{S}=128$ Mbits \cite{E}.

As benchmarks, we consider the optimal cache placement using exhaustive search as well as other benchmarks for all the problems as follows. 
\begin{itemize}
	\item {\bf Greedy caching}: We cache the results of the most energy consuming tasks one by one until the caching capacity is fully exploited. Specifically, we initially set all the tasks as $I_l=0,\forall s_l\in\mathcal{S}$. Then, we solve  $\mathrm{(P0^\prime\text{-}1)}$ to obtain the weighted-sum expected energy consumption. Next, we set the cache placement of the $\bar{l}$th service with the largest energy consumption as $I_{\bar{l}}=1$. Then we repeat the above procedure until constraint \eqref{C:caching capacity constraint} becomes infeasible. This heuristic algorithm is summarized in Algorithm \ref{table:Algorithm 3}.
	\item {\bf Popular caching}: We cache the results of the tasks most likely to be demanded one by one until the caching capacity is achieved. First, we calculate the probability for each service \(s_l\in\mathcal{S}\) to be requested, i.e., $1-\Pr(|\mathcal{K}_l|=0)$, and order these probability in descending order. Next, we cache in descending order the results of those services until the constraint \eqref{C:caching capacity constraint} is violated. 
	\item {\bf No caching}: All task results are not cached, and each task on demand has to be offloaded to and executed at the edge server.
	\item {\bf All caching}: This scheme assumes no constraint \eqref{C:caching capacity constraint}, so all task results are cached at the edge server. It serves as the performance upper-bounds for all other schemes.
	\begin{algorithm}[htp]
		\caption{Greedy Cache Placement Scheme}\label{table:Algorithm 3}
		\SetKwInOut{Input}{Initialize}
		\SetKwInOut{Output}{Output}
		\Input{$\mv I^{(0)}=[0, \cdots,0]^T$, $\mathcal{S}^{(0)}=\mathcal{S}$  and $n=0$}
		\Repeat{\emph{Constraint \eqref{C:caching capacity constraint} becomes infeasible}}{
			Solve $\mathrm{(P0^\prime\text{-}1)}$ to obtain service $s_l$'s expected energy consumption $E_l=\beta_0(\mathbb{E}[E_l^c]+E_l^{\rm dl})+\sum_{k=1}^K\beta_k \mathbb{E}[E_{k,l}^{\rm off}], \forall s_l\in\mathcal{S}^{(n)}$\;		
			Set $I_{\bar{l}}^{(n)}=1$ for service $\bar{l}=\arg\max_{s_l\in\mathcal{S}^{(n)}}E_l$\; 
			Update $\mathcal{S}^{(n+1)}=\mathcal{S}^{(n)} \backslash s_{\bar{l}}$\;
			Update $n=n+1$.	
		}	
		\Output{$\mv I^{(n)}$}
	\end{algorithm}
\end{itemize}
\subsection{DL-Based Offline Cache Placement for $(P0)$} \label{DL}
In the DL-based offline learning framework (c.f.~Fig.~3), the DNN consists of one input layer with 30 neurons, three hidden layers, and one output layer with 10 neurons, where the first, the second and the third hidden layers have 160, 120, and 80 hidden neurons, respectively. Here, we use ReLU as the activation function in the hidden layers and sigmoid activation function in the output layer. We implement the algorithm in Matlab R2020a 9.8 using Deep Learning Toolbox 14.0 and set the learning rate $\eta^{(t)}=0.01$, mini-batch size for training $|\mathcal{D}^{(t)}|=128$, $\forall t$, the data buffer size $|\mathcal{R}|$ as 1024, the training interval $\tau=10$, $M=100$ and $J=10$. We use channel gains and task input/output bits described before to simulate the input data coming of DNN. In addition to the benchmarks described before, we also evaluate the performance of the ``DL-based caching with order-preserving quantization'', in which the order-preserving quantization preserves the ordering of all the entries in a vector during quantization \cite{liang}.

Fig.~5 illustrates the convergence performance of the DL-based cache placement algorithms with different quantization methods using offline implementation. It is observed that both training loss of the DNN with different quantization methods decrease and become stable as time progresses, whose fluctuation is mainly owing to the random sampling of training data. It is worth noting that the algorithm with stochastic quantization method not only wins in training loss, but it is also more robust as the deviation is much smaller. Furthermore, we verify the effectiveness of the trained DNN, whose test loss is also demonstrated in Fig.~5. It is seen that the test loss using stochastic quantization method outperforms the other due to the random exploration of the service caching decisions space. Note that the model inference delay of the proposed framework is mainly dominated by solving problem $\mathrm{(P0^\prime\text{-}1)}$ $J$ times. In the test phase, with $J=5$, the model inference costs around 0.16 s in wall-clock time on average, which is less than 6\% overhead compared with the deadline of $2.8$ s.
\begin{figure}[htp]
	\centering
	\includegraphics[width=3.45in]{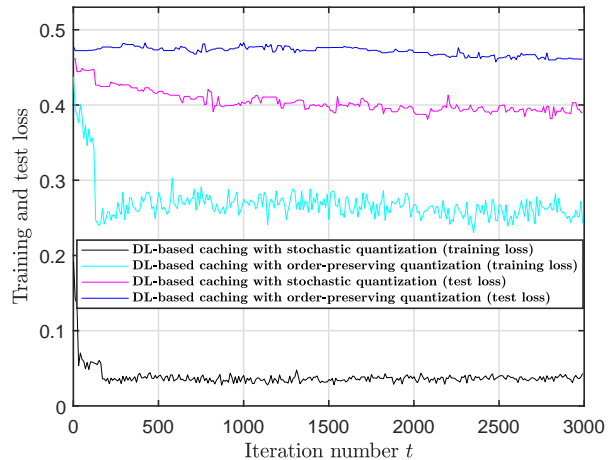}
	\caption{The training and test loss versus the iteration number.}
	\label{loss}
\end{figure}

\begin{figure}[htp]
	\centering
	\includegraphics[width=3.45in]{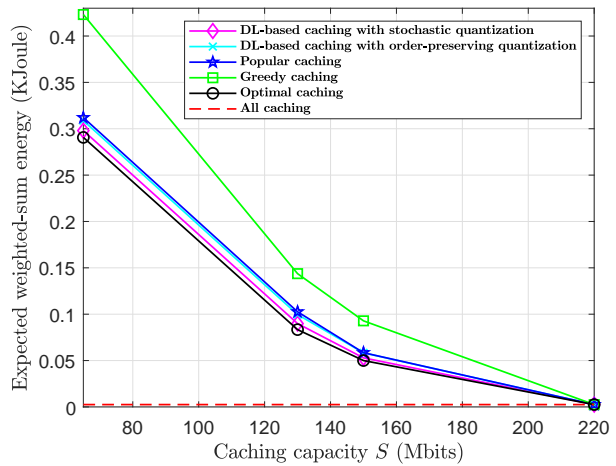}
	\textcolor{black}{\caption{The expected weighted-sum energy versus the caching capacity constraints.}}
	\label{S}
\end{figure}

In Fig.~\ref{S}, we plot the expected weighted-sum energy versus the caching capacity constraint for all caching schemes. It is seen that the expected weighted-sum energy of all schemes drops with the caching capacity. This is intuitively true, as larger caching capacity can accommodate more service results at the edge server. Thanks to the larger diversity brought by the proposed stochastic quantization, the cache placement employing the stochastic quantization outperforms all the other benchmarks, approaching the ``Optimal caching'' when the caching capacity increases. In particular, when the caching capacity exceeds $220$ Mbits, all schemes overlap with the ``All caching'' scheme, since sufficiently large capacity always satisfy $\sum_{s_l\in S}R_l<S$, enabling the trivial case of $I_l^*=1$, $\forall s_l\in S$. In addition, all the shown caching schemes significantly surpass the ``No caching'' one, which yields the expected weighted-sum energy as high as $0.8529$ KJoule.

\begin{figure}[htp]
	\begin{center}
		\resizebox{3.4in}{!}{%
			\begin{tabular}{@{}c|llllllll@{}}
				\toprule
				Deadline $T$ (s) & 2.8 & 2.9 & 3.0 & 3.1 & 3.2 & 3.3 & 3.4 & 3.5  \\ \midrule
				No caching & 0.8529 & 0.6976 & 0.5812 & 0.4918 & 0.4217 & 0.3656 & 0.3201 &0.2825  \\ \bottomrule
			\end{tabular}%
		}
	\end{center}
	\centering
	\includegraphics[width=3.45in]{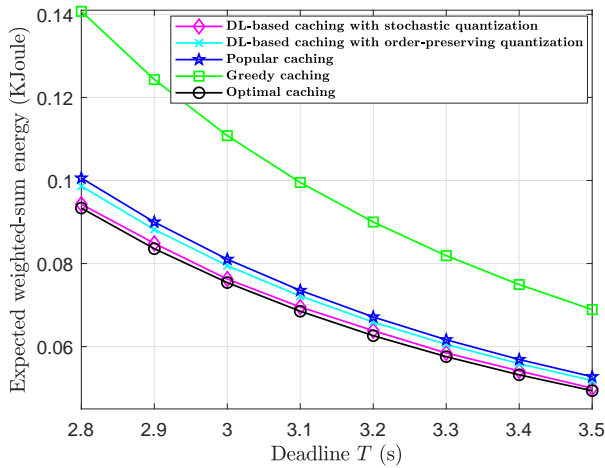}
	\textcolor{black}{\caption{The expected weighted-sum energy versus the deadline constraints.}}
	\label{TE}
\end{figure}

The expected weighted-sum energy versus the computation deadline $T$ for different cache placement schemes is shown in Fig.~\ref{TE}. The weighted-sum energy for all the schemes gradually goes down when the deadline is extended, since more tolerant deadline allows longer execution time for services, thus saving the computation energy $E^{\rm c}$ (c.f.~\eqref{eq:BS computation energy}). In addition, the proposed offline caching with stochastic quantization performs the best among all the suboptimal schemes thanks to the random exploration of the caching capacity, while the one with order-preserving quantization is just slightly better than ``Popular caching'' method. Similar to Fig. \ref{S}, ``No caching'' yields the largest expected weighted-sum energy consumption among all the schemes, which is shown in the table in Fig.~\ref{TE}. 

\begin{figure}[htp]
\centering
\includegraphics[width=3.45in]{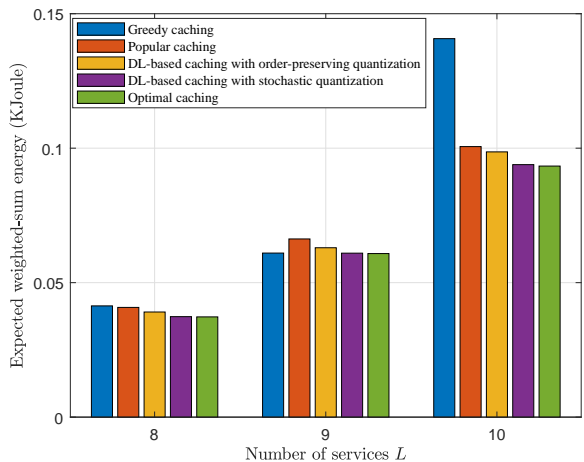}
\textcolor{black}{\caption{The expected weighted-sum energy versus the total number of services.}}
\label{L}
\end{figure} 

Fig.~\ref{L} shows the expected weighted-sum energy consumption for different number $L$ of services with $K=5$ locations. The expected weighted-sum energy consumed by all the schemes increases with the total number of services. The performance gap between the proposed offline caching with stochastic quantization and all the other suboptimal caching schemes enlarges with the number $L$ of services. Specifically, the proposed caching schemes saves $6.12\%$ of energy when there are $L=10$ services versus $1.82\%$ when $L=8$, showing the promising performance of the proposed caching scheme for large $L$. Furthermore, the proposed caching scheme is seen to approach the ``Optimal caching'' with little gap for all values of $L$.    

Fig.~9 demonstrates the expected weighted-sum energy consumption for different weight factors $\beta_0$. It can be seen that the expected energy consumption for all schemes increases with $\beta_0$. This is because the computation energy $E^{\rm c}$ consumed by the BS dominates the energy consumption for delivering a computation service. For example, $E^{\rm c}$ is around 0.171 KJ, $E^{\rm dl}$ is around 0.512$\times 10^{-3}$ KJ, and $E_k^{\rm off }$'s is around 0.475$\times 10^{-4}$ KJ. As a result, the increase in $\beta_0$ will place more weight on $(E^{\rm c}+E^{\rm dl})$, although the weighted-sum energy is minimized. It also shows that our proposed scheme is near-optimal. 

\begin{figure}[htp]
\centering
\includegraphics[width=3.45in]{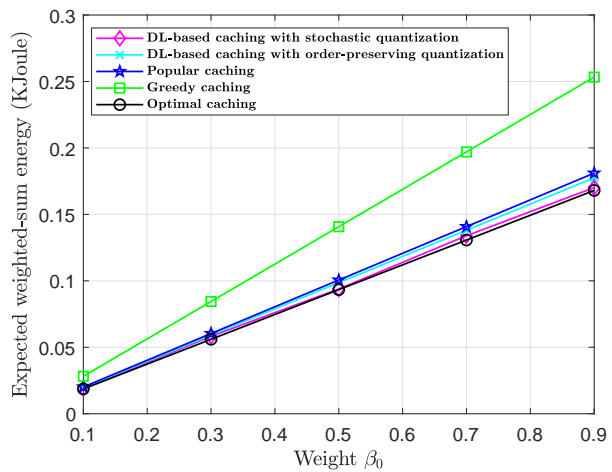}
\textcolor{black}{\caption{The expected weighted-sum energy versus the weight factor $\beta_0$.}}
\label{weight}
\end{figure} 

\subsection{ILP-based Suboptimal Cache Placement for $(P2)$}
In this subsection, we evaluate the performance of the ILP-based caching scheme proposed in section \ref{special} as compared against ``No caching'', ''All caching'' as well as ``Optimal caching''. The parameters considered in this subsection is the same as those in Section \ref{DL}. 
\begin{figure}[htp]
	\centering
	\includegraphics[width=3.45in]{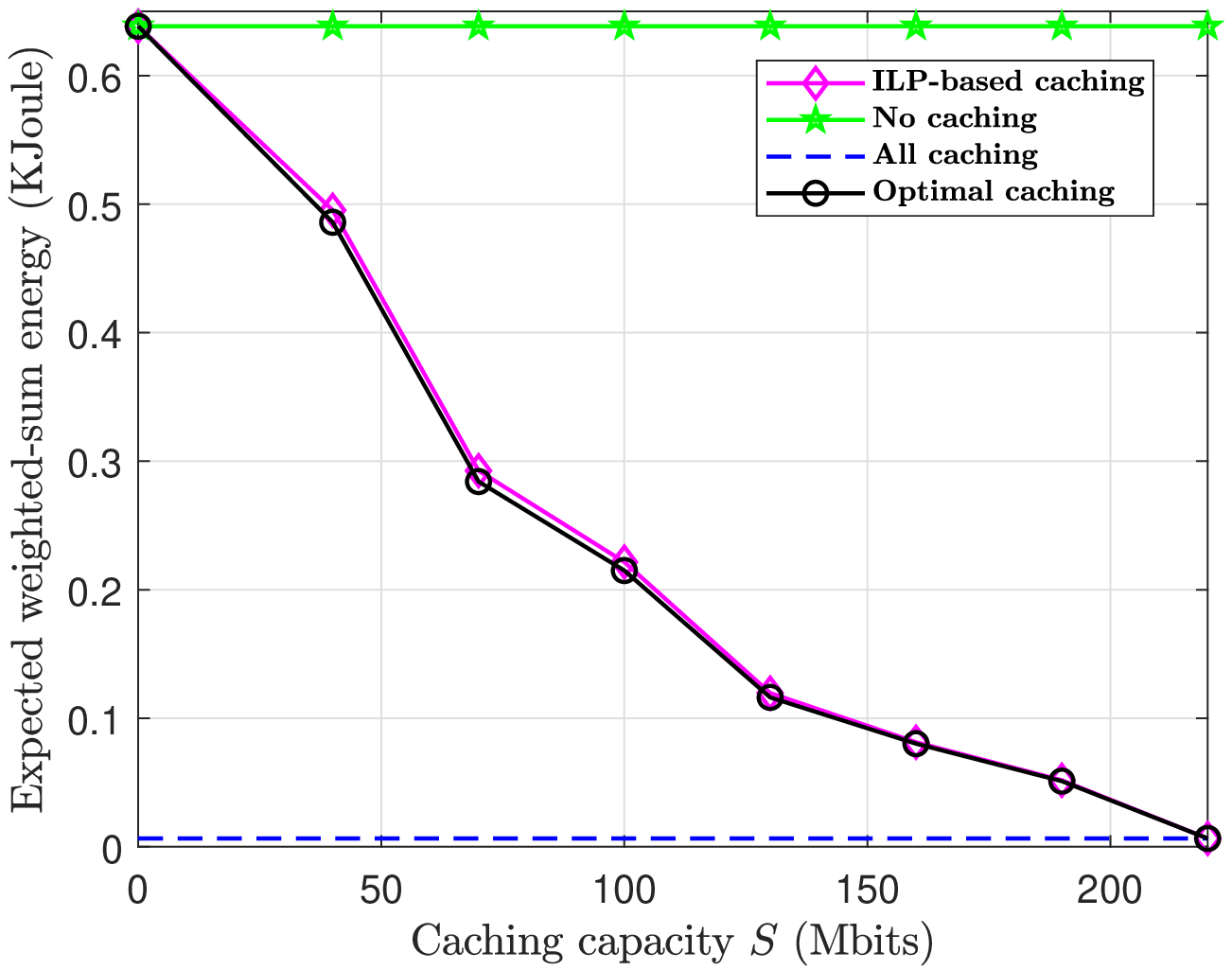}
	\caption{The expected weighted-sum energy versus the caching capacity constraint with $T=3.5$s.}
	\label{Specials}
\end{figure} 
\begin{figure}[htp]
\begin{center}
\resizebox{3.4in}{!}{%
	\begin{tabular}{@{}c|llllllll@{}}
	\toprule
	Deadline $T$ (s) & 3.0 & 3.1 & 3.2 & 3.3 & 3.4 & 3.5  \\ \midrule
	No caching & 1.2674 & 1.0825 & 0.9355 & 0.8166 & 0.7192 &0.6384  \\ \bottomrule
	\end{tabular}%
}
\end{center}
\centering
\includegraphics[width=3.45in]{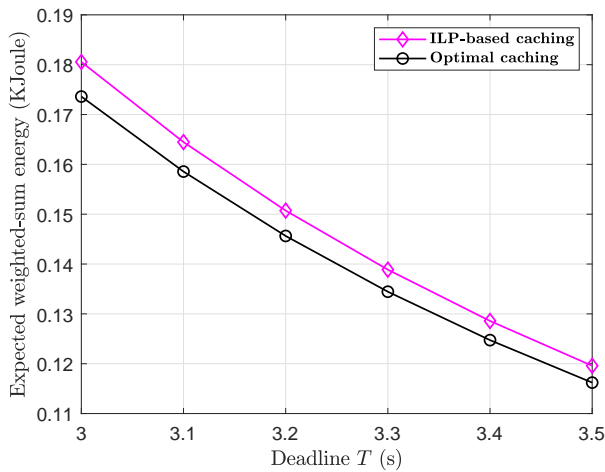}
\caption{The expected weighted-sum energy versus the deadline constraints.}
\label{Special}
\end{figure}

Fig.~\ref{Specials} shows the expected weighted-sum energy versus the caching capacity constraint achieved by different caching schemes. As seen in the general case in Section \ref{DL}, except for the ``No caching'', the weighted-sum energy of all schemes declines with the caching capacity constraint, approaching the all caching scheme with $I_l^\ast=1$, $\forall s_l\in\mathcal{S}$, when $S$ is larger than around $220$ Mbits. Additionally, ``No caching'' is outperformed by all the other caching based schemes as expected.

Next, we demonstrate the expected weighted-sum energy versus the deadline constraints $T$ in Fig.~\ref{Special}. It shows that the ILP-based caching scheme achieves near-optimal performance especially when the deadline constraint $T$ is sufficiently long. This is because longer deadline $T$ allows less computation time, thus leading to lower energy consumption $E^{\rm c}$. In addition, the expected weighted sum energy consumption of the ``All caching'' scheme remains nearly unchanged (0.0064KJoule), since all services have already been cached at the edge server and therefore extending deadline $T$ won't help saving computation energy $E^{\rm c}$ as the other schemes in Fig. \ref{Special}.

\section{Conclusion} \label{sec:Conclusion}
In this paper, we considered a multi-user service-caching enabled MEC system, which serves multiple representative locations with the users at each of them being of a typical preference profile over the given set of computation services. In a FDMA setup, we formulated a joint resource allocation and cache placement optimization problem to minimize the expected weighted-sum energy of the edge server and the users with respect to the location-dependent preference profiles, subjected to the computation, bandwidth and caching capacities as well as the service latency constraints. Under the assumption of known distributions of the channel gains and the task input/output bit-length, we proposed a DL-based service cache placement framework to tackle the mixed-integer challenges, where a DNN is trained offline and then used to predict caching decisions. To achieve better training performance, we also improved  the exploration during training by employing a novel stochastic quantization based caching decision generation scheme. Finally, numerical results showed the striking performance achieved by service caching, in particular, the proposed DL-based service cache placement using stochastic quantization.

Due to space limitation, there are several challenging issues not yet investigated within the scope of this paper, which we summarize here for our future work. First, in this paper, we set the number $J$ of caching decisions as a constant. To avoid severe deviation from optimal solution, $J$ needs to be set relatively large at early phase of the training. When the model is trained for a while, a mild value of $J$ can be set to reduce per-iteration computation complexity. That said, $J$ can be set as a diminishing sequence $\{J^{(t)}\}$ over iterations to gain potentially better training performance \cite{liang}. Moreover, in this paper, we considered a special case of spatial ``correlation'' among tasks, that is, the users at one or multiple locations requesting the same service share the same task-input and task-output data as well as computation requirement. More general cases, where only partial task-input or task-output data corresponding to the same service are overlapped across different locations \cite{2018:x}, or the cached task output can only be partially reused over time horizons \cite{2019:h}, are worthy of further study. At last, we assumed that a cluster of users in one location all have the same channel coefficient in this paper, which is valid in scenarios, e.g., where the users at one location make requests by connecting to a common gateway though, may cause service delay due to the gateway's backlog of requests. Therefore, under the location-representative channel models, detailed formulation accounting for the turn-around time of each local users' request or performance degradation due to inaccurate user-specific channel estimation, will be left for investigations in the future.

\section*{Appendix \uppercase\expandafter{\romannumeral1}\\Proof of lemma of 4.1}
Given a set of (feasible) dual variables, we solve problem \eqref{tc}-\eqref{toff} for their corresponding variables using some of the Karush-Kuhn-Tucker (KKT) conditions \cite{convex} as follows.
\begin{subequations}
\begin{align}
&-2\beta_0\kappa_0(1-\bar{I_l})\frac{(C_lQ_l)^3}{(t_l^c)^3}(1-\prod_{k=1}^K(1-P_{l,k}))+\mu_l-\eta_l=0,\notag\\
&~\forall s_l\in\mathcal{S};
\end{align}
\end{subequations}

\begin{subequations}
\begin{align}
&\begin{cases}
\beta_0p_l^{\rm dl}P_{l,\pi(k)}^{\rm dl}-\gamma_{l,k}\frac{R_l}{(t_{l,\pi(k)}^{\rm dl})^2}=0,\\
~\forall s_l\in\mathcal{S},k\in\mathcal{K}\backslash\{K\},\\
\beta_0p_l^{\rm dl}P_{l,\pi(k)}^{\rm dl}+\mu_l-\gamma_{l,k}\frac{R_l}{(t_{l,\pi(k)}^{\rm dl})^2}=0,\\
~\forall s_l\in\mathcal{S},k=K;
\end{cases} \\
&\begin{cases}
\beta_kp_k^{\rm off}P_{l,k}^{\rm off}-\omega_{l,k}\frac{Q_l}{(t_{l,k}^{\rm off})^2}=0,\\
~\forall s_l\in\mathcal{S},k\in\mathcal{K}\backslash \{K\},\\
\beta_kp_k^{\rm off}P_{l,k}^{\rm off}+\mu_l-\omega_{l,k}\frac{Q_l}{(t_{l,k}^{\rm off})^2}=0,\\
~\forall s_l\in\mathcal{S},k=K.
\end{cases} 
\end{align}
\end{subequations}


After some manipulations, we obtain the optimal solution to \eqref{tc}-\eqref{toff}. Similarly, \eqref{tkdl}-\eqref{alphadls} can also be obtained as above.

\end{document}